\documentclass[letter,11pt]{article}
\pdfoutput=1
\usepackage{jheppub}
\usepackage{hyperref}
\usepackage{graphicx,epstopdf,amsmath,amsfonts,amssymb,appendix,comment} 
\usepackage{color,slashed,subfigure,setspace,footnote,multirow,longtable,braket}
\usepackage{epsfig,mathrsfs,latexsym,color,url,etoolbox}
\usepackage{bbm}
\definecolor{nicered}{rgb}{0.7,0.1,0.1}
\definecolor{nicegreen}{rgb}{0.1,0.5,0.1}
\definecolor{CarnationPink}{rgb}{1.0, 0.65, 0.79}
\DeclareMathAlphabet{\mathbbold}{U}{bbold}{m}{n}    

\usepackage{comment}
\usepackage{mathabx}

\usepackage{float}

\allowdisplaybreaks

\usepackage{dcolumn}
\usepackage{bm}
\usepackage[table]{xcolor}
\usepackage{multirow}
\usepackage{comment}
\usepackage{url}
\usepackage{tcolorbox}
\usepackage[T1]{fontenc}

\definecolor{kjkblue}{rgb}{0.39, 0.589, 0.6914}

\DeclareMathAlphabet{\mathpzc}{OT1}{pzc}{m}{it}

\newcommand{\g}{{\rm g}}
\newcommand{\cm}{{\rm cm}}
\newcommand{\eV}{{\rm eV}}

\newcommand{\MeV}{{\rm MeV}}
\newcommand{\GeV}{{\rm GeV}}

\def\Fermilab{Theoretical Physics Department, Fermilab, P.O. Box 500, Batavia, IL 60510, USA}
\def\Northwestern{Department of Physics \& Astronomy, Northwestern University, Evanston, IL 60208, USA}
\def\COFI{Colegio de F\'isica Fundamental e Interdisciplinaria de las Am\'ericas (COFI), 254 Norzagaray street, San Juan, Puerto Rico 00901}

\graphicspath{{Figs/}}

\begin{document}

\preprint{FERMILAB-PUB-21-459-T, NUHEP-TH/21-15}

\title{
DUNE atmospheric neutrinos: Earth Tomography
}

\author[1]{Kevin J. Kelly,}
\author[1]{Pedro A.~N. Machado,}
\author[1,2,3]{Ivan Martinez-Soler,}
\author[1,2,3]{Yuber F. Perez-Gonzalez}
\affiliation[1]{\Fermilab}
\affiliation[2]{\Northwestern}
\affiliation[3]{\COFI}
\emailAdd{kkelly12@fnal.gov}
\emailAdd{pmachado@fnal.gov}
\emailAdd{ivan.martinezsoler@northwestern.edu}
\emailAdd{yfperezg@northwestern.edu}

\date{\today}

\abstract{
In this paper we show that the DUNE experiment can measure the Earth's density profile by analyzing atmospheric neutrino oscillations.
The crucial feature that enables such measurement is the detailed event reconstruction capability of liquid argon time projection chambers.
This allows for studying the sub-GeV atmospheric neutrino component, which bears a rich oscillation phenomenology, strongly dependent on the matter potential sourced by the Earth.
We provide a pedagogical discussion of the MSW and parametric resonances and their role in measuring the core and mantle densities.
By performing a detailed simulation, accounting for particle reconstruction at DUNE, nuclear physics effects relevant to neutrino-argon interactions and several uncertainties on the atmospheric neutrino flux, we manage to obtain a robust estimate of DUNE's sensitivity to the Earth matter profile.
We find that DUNE can measure the total mass of the Earth at 8.4\% precision with an exposure of 400~kton-year.
By accounting for previous measurements of the total mass and moment of inertia of the Earth, the core, lower mantle and upper mantle densities can be determined with 8.8\%, 13\% and 22\% precision, respectively, for the same exposure.
Finally, for a low exposure run of 60~kton-year, which would correspond to two far detectors running for three years, we have found that the core density could be measured by DUNE at $\sim30\%$ precision.
}

\maketitle
\flushbottom


\section{Introduction}
\label{sec:Introduction}

Understanding the inner structure of the Earth could help us answer many important questions about our planet. 
It is widely believed that the Earth has an is iron-rich core which generates the planet's magnetic field. 
The formation of our planet depends on the core itself, which influences the evolution of the mantle and crust~\cite{Condie}. 
In turn, decays of radio isotopes in the mantle and crust of the Earth, in particular $^{40}$K, $^{232}$Th and $^{238}$U, originate a large flux of so-called geoneutrinos~\cite{KamLAND:2011ayp, KamLAND:2013rgu, Borexino:2019gps}. 
which are crucial in the cooling mechanism of our planet~\cite{davies2010earth, McDonough:2019ldt}.
Nevertheless, understanding the Earth's matter profile remains an extraordinary endeavor.

To understand the matter profile, we need to rely on indirect measurements.
The most traditional of those measurements are performed with seismological data.
By studying seismic waves in the surface of the planet, we can infer what is the matter profile that these waves went through.
Combining observations at different positions at the surface and from many earthquakes allows geologists to measure the Earth's density as a function of the depth.
While these measurements provide an excellent model of the matter distribution, they rely on empirical relations that describe how compressional and transverse waves, typically referred to as P (primary) and S (secondary) waves, propagate on a dense medium (see e.g. Ref.~\cite{Geller:2001ix}), as well as assumptions on the variation of the matter profile as a function of the depth.
On top of that, a clearcut statistical evaluation of the uncertainties on the density profiles in these measurementsis not straightforward~\cite{Kenneth1998}.
Thus, it would be desirable to measure the Earth's matter profile with alternative methods.

Indeed, there have been several proposals to use atmospheric neutrinos, which originate from the decays of hadrons created when cosmic rays hit the atmosphere, to measure the matter profile.
The most conceptually simple method is via the attenuation of atmospheric neutrinos~\cite{DeRujula:1983ya, Wilson:1983an, Askarian:1984xrv, Borisov:1986sm, Jain:1999kp, Reynoso:2004dt, Gonzalez-Garcia:2007wfs, Agarwalla:2012uj, Ioannisian:2017chl, Donini:2018tsg}.
Atmospheric neutrinos span several orders of magnitude in energy, from the tens of MeV to beyond tens of TeV.
While neutrinos interact weakly with matter, the probability for a multi-TeV neutrino to cross the Earth without interactions is significantly different from unity.
The mean free path of TeV neutrinos crossing the Earth can be written as
\begin{equation}
  \frac{\lambda}{R_{\scriptscriptstyle\oplus}} \sim 3.7 \left(\frac{10~\text{g/cm}^3}{\rho}\right)\left(\frac{10~\text{TeV}}{E}\right),
\end{equation}
where $R_{\scriptscriptstyle\oplus}\simeq6371$~km is the radius of the Earth and $E$ is the energy of the neutrino.
Thus, by analyzing the attenuation of high energy atmospheric neutrinos as a function of the zenith angle of the incoming neutrinos, the matter profile of the Earth may be inferred (as in, e.g., Ref.~\cite{Donini:2018tsg}).

As we will see later, the atmospheric neutrino flux drops sharply with larger energies: a much larger flux of neutrinos is then available at lower energies, particularly below the GeV scale.
If we calculate the mean free path of neutrinos below ${\sim}$TeV, though, we will find that atmospheric flux attenuation becomes small to the point that it is experimentally unobservable.
This brings us to the other way of measuring the Earth's interior: probing the Earth's density with neutrino oscillations~\cite{Nicolaidis:1990jm, Ohlsson:1999um, Lindner:2002wm, Akhmedov:2005yt, Winter:2006vg, Rott:2015kwa, Winter:2015zwx, DOlivo:2020ssf, Kumar:2021faw}.

The oscillation of atmospheric neutrinos was first measured by the Kamiokande experiment and, together with the observation of solar neutrinos by SNO, it led to the discovery of nonzero neutrino masses~\cite{McDonald:2016ixn, Kajita:2016cak}.
Neutrino oscillation is, in fact, a simple and yet exquisite quantum mechanical effect. 
Charged-current weak interactions couple charged leptons to a superposition of neutrino states with well defined masses. 
More precisely,  
\begin{equation}
  |\nu_\alpha \rangle = \sum_i U_{\alpha i}^* |\nu_i \rangle,
\end{equation}
where $|\nu_\alpha \rangle$ ($\alpha = e,\ \mu,\ \tau$) denotes the flavor eigenstates which couple diagonally to each flavor of charged lepton and the $W$ boson, and $|\nu_i \rangle$ are mass eigenstates which have well-defined masses.
The $3\times3$, unitary, Pontecorvo-Maki-Nakagawa-Sakata (PMNS) matrix $U_{\alpha i}$ parametrizes the mixing among flavor and mass eigenstates.

In a simplified two neutrino flavor oscillation scheme, the transition probability $\nu_\alpha\to\nu_\beta$, for different flavors, can be written as
\begin{equation}
  P_{\alpha \beta} = \sin^2(2\theta)\sin^2\left(\frac{\Delta m^2 L}{4E}\right),\qquad \alpha\neq\beta,
\end{equation}
where $\Delta m^2 \equiv m_2^2-m_1^2$ is the squared neutrino mass splitting, $\theta$ is the angle that parametrizes the mixing and $L$ is the distance traveled by the neutrino, or its baseline. 
While in reality we know that there are three neutrinos, the above formula is still a useful guide.
This is because the so-called ``atmospheric splitting''  $|\Delta m^2_{31}|\sim 2.5\times10^{-3}$~eV$^2$~\cite{DayaBay:2018yms, NOvA:2019cyt, T2K:2021xwb} is about 30 times larger than the ``solar splitting'' $\Delta m^2_{21}\sim7.5\times10^{-5}$~eV$^2$~\cite{KamLAND:2013rgu, yasuhiro_nakajima_2020_4134680}.
Oscillation effects stemming from these two oscillation frequencies, to first order, decouple from each other.

A key point of neutrino oscillations is that usual matter sources a potential that affects neutrino propagation, changing their dispersion relation in a flavor dependent way and acting similarly to a refraction index~\cite{Wolfenstein:1977ue, Mikheyev:1985zog, Mikheev:1986wj}.
Observing how atmospheric neutrinos oscillate as they cross the Earth at different zenith angles can, in principle, allow us to infer the matter profile of the Earth through a genuine quantum mechanical effect.
Moreover, there is an interesting  complementarity among seismology, neutrino absorption, and neutrino oscillation measurements of the matter profile.

The propagation of seismic waves depend on the total density of matter, as well  as on variations of this density and on the state of matter (solid vs. liquid) through the wave trajectory.
In particular the waves that oscillate perpendicular to the direction of propagation (S-waves) do not travel through fluids.
The absence of S-waves detected in a surface region diametrically opposite to an earthquake location is a strong indication that the core of our planet is liquid, besides providing a robust measurement of the liquid core boundaries.
In contrast, absorption of high energy atmospheric neutrinos depends mainly on the number of the proton and neutrons in the incoming neutrino line-of-sight, regardless of how the density profile varies.
To obtain the density profile, one needs to be able to measure the incoming neutrino direction and reconstruct the density profile by analyzing the  zenith-dependent neutrino attenuation.
The component of the neutrino matter potential that affects neutrino oscillation, on the other hand, is sourced only by electrons and it is sensitive to the number density distribution along the neutrino line of sight, as we will discuss in detail later.
Therefore, since the Earth is electrically neutral, combining seismology, neutrino absorption and neutrino oscillations could in principle allow for a measurement of the average neutron to proton ratio inside the Earth, at different radii, which would ultimately teach us about the average chemical composition of Earth's core.

In this manuscript, we are interested in how to measure the Earth's density profile with atmospheric neutrino oscillations. 
There are two energy regions of interest here.
While oscillations driven by the atmospheric splitting $\Delta m^2_{31}$ are dominant in the few-10~GeV region, those driven by $\Delta m^2_{21}$ are more relevant around 0.1-1~GeV.
The latter not only benefit from a higher neutrino flux, but also a larger oscillation amplitude due to larger mixing angles.
Because of that, we will pay particular attention to the sub-GeV atmospheric neutrino component, even if we include atmospheric neutrinos from 0.1 to 100 GeV in our analysis.

Due to the small interaction cross section, only multi-kiloton detectors are capable of measuring atmospheric neutrinos with sufficient statistics.
To date, the only multi-kiloton detectors available are Cherenkov observatories, like Super-Kamiokande and IceCube. 
The energy and direction of charged particles traversing these observatories, as well as their type (e.g. electrons versus muons) can be inferred  by  detecting the Cherenkov light emitted by these particles.
This allows to reconstruct the incoming neutrino energy, direction and flavor, which is crucial to extract physics from atmospheric neutrinos.
Nevertheless, particles only radiate Cherenkov light if they are faster than light in that medium. 
In particular, protons with total energies below 1.4~GeV do not emit Cherenkov light in water.
Because of that, it is nearly impossible for Cherenkov detectors to leverage the atmospheric neutrino flux below the GeV scale, as the correlation between the outgoing lepton and the incoming neutrino directions is very weak. 
This lack of directionality for sub-GeV neutrinos reduces the sensitivity of Cherenkov detectors  to the Earth density profile. 
Both Super-Kamiokande~\cite{Super-Kamiokande:2017yvm} and IceCube~\cite{IceCube:2019dyb} collaborations have performed analyses on the sensitivity to the overall matter potential sourced by the Earth, indicating preference for a nonzero value.

This bring us to the Deep Underground Neutrino Experiment (DUNE), a 40~kton liquid argon time projection chamber being built below the Black Hills of South Dakota~\cite{DUNE:2020lwj, DUNE:2020ypp}. 
Liquid argon time projection chambers, or LArTPCs for short, detect the ionization energy of charged particles traversing the argon, which enables the identification and tracking of particles, including low energy ones.
In fact, the ArgoNeuT experiment at Fermilab has shown that it is possible for those detectors to reconstruct protons with as low as 21~MeV of kinetic energy~\cite{ArgoNeuT:2018tvi}.

This immediately prompts us to investigate how much physics DUNE could extract from the sub-GeV atmospheric neutrino sample.
Although the large flux and detector reconstruction capabilities are very promising, there are inherent difficulties in studying sub-GeV atmospheric neutrinos  at DUNE.
Modeling neutrino-nucleus interactions at the 0.1-1~GeV region is remarkably challenging~\cite{Benhar:2015wva}.
First, there are difficulties associated to the hard interaction itself, due to Fermi motion of the nucleons inside the nucleus and nuclear binding energy~\cite{Dutta:2000sn, JLabE91013:2003gdp, Benhar:1994hw, Dickhoff:2004xx, Benhar:2006wy, Ankowski:2014yfa, Benhar:2015ula, Rocco:2015cil, Barbieri:2016uib, Rocco:2018mwt, Rocco:2020jlx}, in-medium effects that may change the local dispersion relation of those nucleons~\cite{Arnold:1981dt, Hama:1990vr, Cooper:1993nx}, and even nucleon-nucleon correlations in the nuclear medium~\cite{Mathiot:1980js, Kohno:1981dg, Dehesa:1985asb, Jiang:1992wn, Sobczyk:2012ms, Megias:2014qva, Megias:2016fjk}.
Then, even if these obstacles are overcome, one still needs to properly model how nucleons propagate within the nucleus, including how the nuclear potential and intranuclear cascades affect the energy, angular and multiplicity distributions of final state nucleons~\cite{Bertini:1963zzc, Cugnon:1980zz, Bertsch:1984gb, Bertsch:1988ik, Cugnon:1996xf, Boudard:2002yn, Sawada:2012hk, Uozumi:2012fm, Golan:2012wx, Isaacson:2020wlx, Dytman:2021ohr}.

All these effects are important because, in order to extract useful information from sub-GeV atmospheric neutrinos, it is crucial to reconstruct the incoming neutrino direction~\cite{Kelly:2019itm}.
To do so, the energy, direction and type of outgoing particles should be reconstructed and combined to obtain the incoming neutrino four momentum in an event-by-event basis.
Most of the aforementioned effects, especially intranuclear cascades, will change the direction and multiplicity of outgoing protons and neutrons, the former being essentially invisible at LArTPCs.

Despite all these challenges, previous work has shown that the sub-GeV sample alone can yield valuable information on $CP$ violation~\cite{Kelly:2019itm}.
Here we perform a detailed study of DUNE's capability to yield the first quantum tomography measurement of the Earth's matter profile using neutrino oscillations.
We consider various systematic uncertainties on the atmospheric neutrino flux, as well as state-of-the-art neutrino-nucleus interactions using the \texttt{NuWro} neutrino-nucleus event generator~\cite{Golan:2012rfa}, and realistic detector responses to particle identification and reconstruction.
We will show that the future DUNE experiment has the capability of pioneering a quantum tomography measurement of our planet, determining the core density at the 10\% level.

This paper is organized as follows.
In Sec.~\ref{sec:AtmoNu} we discuss neutrino oscillations in the Earth, focusing on the effects induced by matter; in Sec.~\ref{sec:DUNEDetails} we describe the DUNE experiments and how we model the detector response; in Sec.~\ref{sec:FluxUncertainties} the atmospheric neutrino flux and associated uncertainties are discussed; in Sec.~\ref{sec:ResultsDiscussion} we present our results; and finally we conclude in Sec.~\ref{sec:Conclusions}.
We use natural units where $\hbar = c = k_{\rm B} = 1$ throughout this manuscript, unless otherwise stated.

\section{Oscillations of Atmospheric Neutrinos}
\label{sec:AtmoNu}

In this section, we review the physics of atmospheric neutrino oscillations.
Before laying out the oscillation formalism and elaborating on the effects of nonzero matter density, we first discuss the Earth density profile.

\subsection{Earth Density Profile}

The current knowledge of the Earth density profile comes from seismological studies, summarized in the \emph{preliminary Earth reference model}~\cite{Dziewonski:1981xy}\footnote{More recent, three dimensional models of the Earth matter profile exist~\cite{REM}, but we will nonetheless avoid dealing with asymmetric profiles.}.
Measurements of the precise matter density and variations of it with respect to the depth are obtained using empirical relations for the wave velocities~\cite{Geller:2001ix}.
The determination of the radii separating the core, inner mantle and outer mantle, on the other hand, is a much more robust observation, as it relies on the reflection of waves.
In view of that, our study will assume known radii separating the different layers of the Earth, while trying to constrain the matter densities in the layers.
Moreover, in many examples and analysis we will keep the mass of the Earth and its moment of inertia fixed, as these quantities are extremely well measured~\cite{ries1992progress, Rosi:2014kva, williams1994contributions, chen2015consistent}.
We will discuss these in details in Sec.~\ref{sec:ResultsDiscussion}.

Since neutrino oscillation data will not have the same fine precision as seismological data, we adopt a simplified model of the Earth profile with three layers\footnote{Since the neutrino trajectory through the crust is a small fraction of its overall path, we do not include the crust in our simulations.}: the core with density $\rho_{\rm C}=11$~g/cm$^3$, the lower mantle with density $\rho_{\rm LM}=5.1$~g/cm$^3$; and upper mantle with $\rho_{\rm UM}=3.1$~g/cm$^3$. As a function of the radius from the center of the Earth, these layers correspond to $r < 3480$ km (core), $3480$ km $\leq r < 5700$ km (lower mantle), and $5700 < r < R_{\scriptscriptstyle\oplus} = 6371$ km (upper mantle).

The possible neutrino trajectories through the Earth depending on the zenith angle $\zeta$ are the following.
\begin{enumerate}
	\item \emph{Core-Mantle}: Neutrinos whose zenith angle is $\zeta \gtrsim 147^{\circ}$ cross the three main layers, core, lower and upper mantle. In this case, the relatively large difference between the densities of the core and mantle  can lead to a parametric resonance, which will be discussed later, depending on the neutrino energies. The specific values for zenith angles and energies where such parametric amplification will depend on the actual values of the densities. 
	\item \emph{Upper-Lower Mantle}: For $116^{\circ} \lesssim \zeta \lesssim 147^{\circ}$, neutrinos only cross the upper and lower mantle layers. We will see that from the known values of the densities, neutrino oscillations can be resonantly enhanced because of the MSW effect for some specific values of the energies and zeniths.
	\item \emph{Upper Mantle Only}: Finally, if neutrinos have a $\zeta \lesssim 116^{\circ}$, they only propagate in the Upper mantle. Neutrino oscillations are then well described by considering matter with constant density. 
\end{enumerate}

\subsection{MSW and Parametric Resonances}

In this section we will review the oscillation phenomenology of atmospheric neutrinos. 
First, we remind the reader that, in the presence of matter, the evolution operator that drives neutrino oscillations is given in the flavor basis by
\begin{equation}\label{eq:hamiltonian}
  H = \frac{1}{2E}
U^\dagger\left(
\begin{array}{ccc}
 0 & 0  & 0  \\
 0 & \Delta m^2_{21}  &  0  \\
 0 & 0  &  \Delta m^2_{31} 
\end{array}
\right)U
+
\left(
\begin{array}{ccc}
 V_{\rm CC} & 0  & 0  \\
 0 & 0  &  0  \\
 0 & 0  &  0 
\end{array}
\right),
\end{equation}
where $U$ is the PMNS matrix, $E$ is the neutrino energy, $\Delta m^2_{ij}\equiv m_i^2-m_j^2$ denotes the solar $\Delta m^2_{21}$ and atmospheric $\Delta m^2_{31}$ mass splittings; and 
\begin{equation}
  V_{\rm CC} = \sqrt{2} G_F n_e
\end{equation}
is the matter potential sourced by electrons, with $G_F$ being the Fermi constant and $n_e$ the electron number density. 
Although electrons, protons and neutrons also induce a neutral current matter potential for neutrinos, the assumption of electricaly neutral matter cancel the contribution of the charged particles. The contribution from neutrons is flavor universal and does not change oscillation phenomenology.
When $V_{\rm CC}$, or equivalently $n_e$, is constant along the path of the neutrino, the evolution of a neutrino state after is propagates a baseline $L$ is simply obtained by $|\nu(L)\rangle = e^{-iHL} |\nu(0)\rangle$. 
The matter potential plays a crucial role in the oscillation of atmospheric neutrinos.
In the following, we will discuss how the Earth density profile affects neutrino oscillations, emphasizing the MSW and parametric resonances.

Neutrino flavor oscillations inside the Earth can present a unique type of resonant enhancement, the \emph{parametric resonance}, due to the abrupt change in density existing between terrestrial layers.
Such amplification results from serendipitous relations between the oscillation phases in the different layers that a neutrino can traverse~\cite{Akhmedov:1998ui,Akhmedov:1998xq,Chizhov:1998ug,Chizhov:1999az,Chizhov:1999he,Liu:1998nb}.
Crucially, the specific relations that the phases need to fulfill depend on the matter profile, so any modification of the densities will lead to significant changes on the oscillations observed in a detector. 

\textbf{MSW Resonances:} While we know that there are three neutrinos in nature, the two neutrino system will provide us with a simple yet useful framework to understand the key features of atmospheric neutrino oscillations.
When neutrinos propagate inside matter, flavor oscillations are modified due to the presence of the potential associated to the electron density. 
When matter sources a potential, the Hamiltonian \eqref{eq:hamiltonian} is no longer diagonalized by the PMNS matrix $U$, which in the two neutrino framework is just a rotation matrix with an angle $\theta$.
Instead, the diagonalization is obtained by an effective mixing angle in matter $\widetilde\theta$ and yields eigenvalues   $\Delta\tilde m^2/2E$, where the numerator denotes the effective mass splitting in matter.
It is easy to show that
\begin{align}
	\Delta\widetilde m^2=\sqrt{\left(\Delta m^2 \cos 2\theta - A_{\rm CC}\right)^2+\left(\Delta m^2 \sin 2\theta\right)^2},\quad \sin 2\widetilde\theta =\frac{\Delta m^2 \sin 2\theta}{\Delta\widetilde m^2},
\end{align}
which depend on the vacuum mixing and mass splitting ($\Delta m^2$), as well as and on the matter contribution $A_{\rm CC}$.
This matter contribution is defined by
\begin{align}
 	A_{\rm CC} = 2\sqrt{2}G_{\rm F} N_A\, E\, Y_e\, \rho \approx 4.5 \times 10^{-4}\ \mathrm{eV}^2 \left(\frac{E}{\mathrm{GeV}}\right) \left(\frac{Y_e}{0.5}\right) \left(\frac{\rho}{6\ \mathrm{g/cm}^3}\right),
\end{align}
where we have written the electron number density as $n_e = N_A Y_e \rho$, in which $N_A$ the Avogadro constant, $Y_e$ the electron fraction, and $\rho$ the density of the medium. 

We see that the presence of matter can modify the mixing angle, and in particular can yield $\sin2\tilde \theta=1$.
This happens at a neutrino energy given by
\begin{equation}\label{eq:energy_msw}
	E_{\rm MSW}=\frac{\Delta m^2 \cos 2\theta}{2\sqrt{2}G_{\rm F}N_AY_e\rho} \simeq
		\begin{cases}
		 5.3~\GeV \left(\frac{\Delta m^2}{2.5\times10^{-3}~\eV^2}\right)\left(\frac{\cos 2\theta}{0.95}\right)\left(\frac{0.5}{Y_e}\right)\left(\frac{6~\g/\cm^{3}}{\rho}\right) &\quad \text{(atmospheric pars.)}\\
		 68~\MeV \left(\frac{\Delta m^2}{7.5\times10^{-5}~\eV^2}\right)\left(\frac{\cos 2\theta}{0.41}\right)\left(\frac{0.5}{Y_e}\right)\left(\frac{6~\g/\cm^{3}}{\rho}\right) &\quad \text{(solar pars.)}
		\end{cases}
\end{equation}
where we took as reference values the atmospheric mass splitting $\Delta m^2_{31}$ and $\theta_{13}$ in the upper line and the solar splitting and $\theta_{12}$ for the lower line, as well as a representative matter density.
We see that even a small vacuum mixing angle can be enhanced by matter effects to drive large flavor transitions.
The transition probability itself is simply
\begin{align}
	P_{\alpha\beta}^{\rm MSW}= \sin^22\widetilde\theta\sin^2\left(\frac{\Delta\widetilde m^2 L}{4E}\right),\qquad \alpha\neq\beta.
\end{align}
A maximal transition probability, $P_{\alpha\beta}^{\rm MSW}=1$, will be possible for a specific neutrino energy, see Eq.~\eqref{eq:energy_msw}, and when the \emph{phase} of oscillation $\phi$ has the right value,
\begin{align*}
	\phi\equiv\frac{\Delta\widetilde m^2 L}{4E} = \frac{2k+1}{2}\pi,\quad k=0,1,2,\ldots.
\end{align*}
Thus, to have a complete flavor conversion in matter, both the MSW energy and phase conditions should be fulfilled simultaneously. 

Assuredly, the accomplishment of such conditions takes place for very specific values of the neutrino energy and distance. 
However, atmospheric neutrinos cover a broad energy spectrum, from the tens of MeV up to multi-TeV, as well as a large baseline interval, from hundreds of kilometers to the Earth's diameter of $12,742$~km.
Hence, one would hope that the maximal transition probability can be observed for atmospheric neutrinos if the incoming neutrino energy and direction can be reconstructed with some  accuracy.

To get a more quantitative idea, if we take the average Earth density $\bar{\rho} = 5.5 {\rm~g/cm^{3}}$ as a representative example, we find that the MSW resonance is achieved for $E_{\rm MSW}\simeq 5.73~\GeV$  for the atmospheric splitting, while  $E_{\rm MSW}\simeq 70~\MeV$ for the solar sector.
Meanwhile, the smallest distance required to have a maximal transition should be $L=9,\!580~{\rm km}$ and $L=1,\!258~{\rm km}$ for the atmospheric and solar sectors, which correspond to zenith angles of $\zeta = 138^{\circ}$ and $\zeta = 95^{\circ}$, respectively.
While this suggests that two MSW resonances indeed take place for atmospheric neutrinos, this is not the whole story: the multi-layer structure of the Earth significantly modifies the neutrino state evolution, leading to parametric resonances.

\textbf{Parametric Resonances:} The description of neutrinos propagating in a matter profile consisting of layers with different densities has been studied extensively~\cite{Akhmedov:1998ui,Akhmedov:1998xq,Chizhov:1998ug,Chizhov:1999az,Chizhov:1999he,Liu:1998nb,Peres:1999yi,Freund:1999vc,Palomares-Ruiz:2004cmm,Akhmedov:2005yj,Akhmedov:2006hb,Gandhi:2004bj,Gandhi:2004md,Kimura:2004vh,Gonzalez-Garcia:2004pfd,Liao:2007re,Akhmedov:2008qt,Jacobsson:2001zk,Ohlsson:1999um,Koike:2009xf,Akhmedov:2016hcb}.
For completeness, we provide here a brief discussion and derivation of the required conditions that lead to a parametric enhancement of the oscillation of atmospheric neutrinos. First, let us consider the evolution operator for two-neutrino oscillations in the flavor basis. This operator evolves an initial state $\nu(L_0)$ to a final state $\nu(L)={\cal S}(L-L_0)\nu(L_0)$, where the two-flavor state is $\nu^{2\times 2}=(\nu_\alpha, \nu_\beta)^T$. As above, for a time-invariant Hamiltonian, ${\cal S}(L)\equiv {\rm exp}(-iHL)$. To study parametric resonances, we will first assume that the Earth consists of two layers, the core and the mantle, each with a constant density $\rho_{\rm C}$ and $\rho_{\rm M}$, respectively.

The $2\times2$ Hamiltonian can be written as a linear combination of Pauli matrices, since any component proportional to the identity matrix does not contribute to neutrino oscillations. Thus we can always write the evolution operator ${\cal S}_X$ for a layer $X = \mathrm{C,\ M}$ as~\cite{Akhmedov:1998ui,Akhmedov:1998xq,Chizhov:1998ug,Chizhov:1999az,Chizhov:1999he}
\begin{align}
	{\cal S}_X = \cos \phi_X \mathbbm{1}_2- i\sin \phi_X \, \vec{\sigma}\cdot\overrightarrow{n_X},
\end{align}
where  $\vec{\sigma}\equiv\{\sigma_1,\sigma_2,\sigma_3\}$ is a vector of Pauli matrices and $\mathbbm{1}_2$ is the $2\times 2$ identity matrix.
The oscillation phase in the layer is given by
\begin{align*}
\phi_X = \frac{\Delta\widetilde m_{X}^2 L_X}{4E}
\end{align*}
being $L_X$ the distance traveled and $\Delta\widetilde m_{X}^2$ the effective mass splitting  within the layer. 
The unit vector $ \overrightarrow{n_X}$ corresponds to
\begin{align}
	\overrightarrow{n_X} = (\sin 2\theta_X, 0, -\cos 2 \theta_X),
\end{align}
where $\theta_X$ is the angle in matter within the layer $X$.
Note that we have dropped the tilde in the effective mixing angles in matter to avoid clutter, as the layer subscript already distinguishes them from the vacuum angle.
This parametrization of the evolution operator is useful to understand the evolution of neutrinos inside the Earth. 

\begin{figure}[t]
\begin{center}
\includegraphics[width=0.55\linewidth]{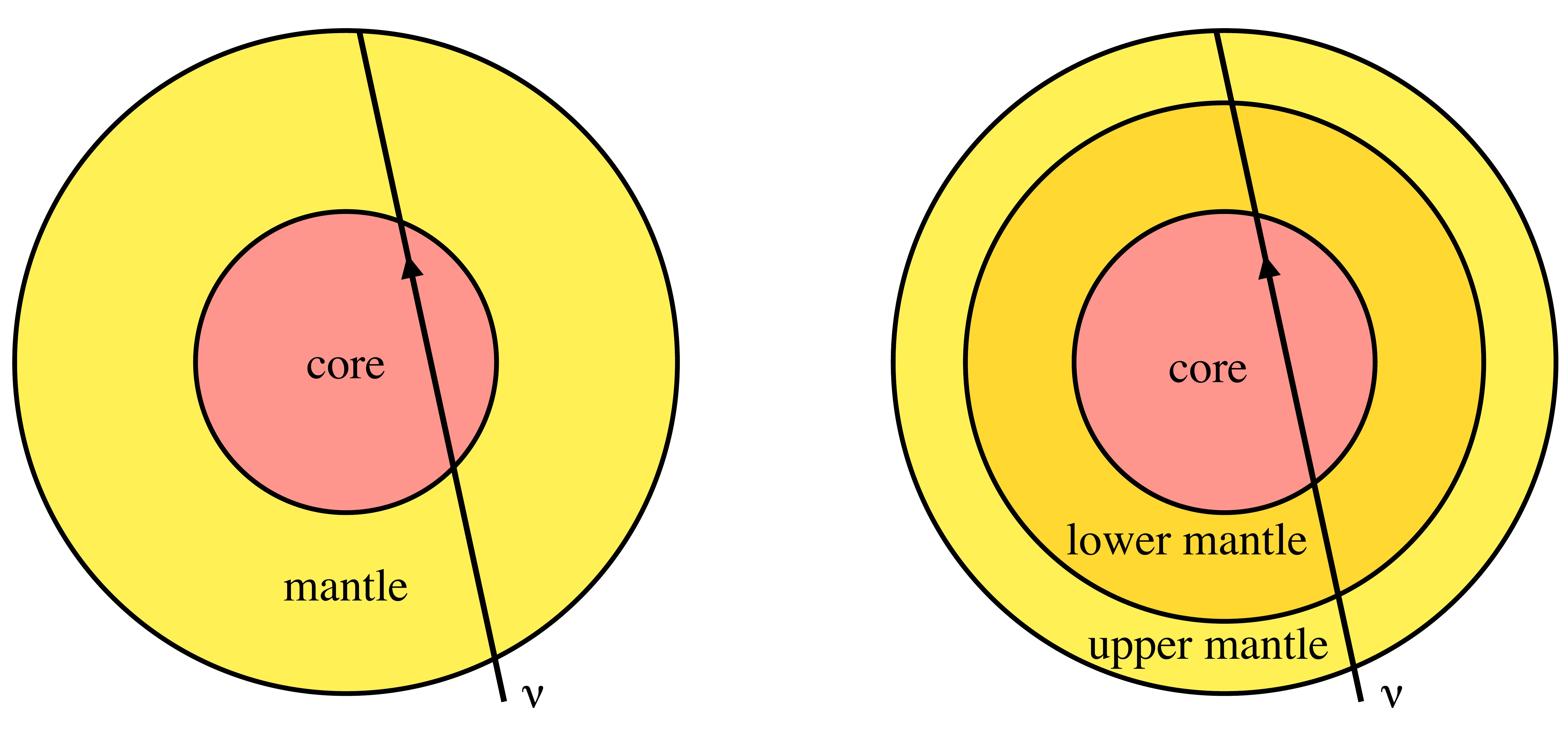}
\caption{Pictorial representation of toy 2-layer Earth model (left) and 3-layer model (right). \label{fig:layers}} 
\end{center}
\end{figure}
To begin with, let us consider neutrinos crossing two different layers, that is, neutrino go through the mantle, then through the core and back to the mantle, as pictorially shown in the left panel of Fig.~\ref{fig:layers}. 
In this case, the full evolution operator consists in the multiplication of operators for each crossed layer,
\begin{align*}
	{\cal S} &= {\cal S}_{\rm M}\,{\cal S}_{\rm C}\,{\cal S}_{\rm M},
\end{align*}
Such multiplication can be presented in a simpler form after using the properties of Pauli matrices,
\begin{align}
	{\cal S} & = W_0 \mathbbm{1}_2 - i \vec{\sigma}\cdot\overrightarrow{W},
\end{align}
where $W_0, \overrightarrow{W}=\{W_1,0,W_3\}$, depend on the phases and amplitudes in both layers~\cite{Akhmedov:1998ui,Akhmedov:1998xq,Chizhov:1998ug,Chizhov:1999az,Chizhov:1999he}, namely
\begin{subequations}
\begin{align}
 W_0 &= \cos 2\phi_{\rm M}\cos \phi_{\rm C} -\cos 2(\theta_{\rm M}-\theta_{\rm C}) \sin 2\phi_{\rm M}\sin \phi_{\rm C},\\
 W_1 &=  \sin 2 \theta_{\rm M}\sin 2\phi_{\rm M}\cos \phi_{\rm C} +(\cos 2(\theta_{\rm M}-\theta_{\rm C}) \sin 2\theta_{\rm M}\cos 2\phi_{\rm M}- \cos2\theta_{\rm M}\sin 2(\theta_{\rm M}-\theta_{\rm C}))\sin \phi_{\rm C},\\
 W_3 & = -\sin 2 \theta_{\rm M}\sin 2(\theta_{\rm M}-\theta_{\rm C}) \sin\phi_{\rm C} - (\sin2\phi_{\rm M}\cos\phi_{\rm C}+\cos 2(\theta_{\rm M}-\theta_{\rm C}) \cos 2 \phi_{\rm M}  \sin\phi_{\rm C})\cos 2\theta_{\rm M}.
 \end{align}
\end{subequations}
Although complicated at first sight, these contain the conditions for a parametric amplification of oscillations. 
Notice that the two-neutrino survival and appearance probabilities are simply
\begin{align}
	P_{\alpha\alpha}^{\rm 2f} = |W_0|^2+|W_3|^2; \hspace{3cm}
  	P_{\alpha\beta}^{\rm 2f}  = |W_1|^2\quad (\alpha\neq\beta).
\end{align}

A complete neutrino flavor conversion can only take place if
\begin{align}
	W_0 = 0,\qquad W_3=0.
\end{align}  
The solution of the previous system of equations is given by~\cite{Chizhov:1998ug,Chizhov:1999az,Chizhov:1999he},
\begin{align}\label{eq:SolPR}
	\tan^2\phi_{\rm M}&=-\frac{\cos 2\theta_{\rm C}}{\cos 2(2\theta_{\rm M}-\theta_{\rm C})},\quad\text{and}\quad\tan^2\phi_{\rm C}=-\frac{\cos^2 2\theta_{\rm M}}{\cos 2\theta_{\rm C}\cos 2(2\theta_{\rm M}-\theta_{\rm C})}.
\end{align}
These conditions relate the phases and mixing angles in matter between the two layers: the oscillation phases in both layers depend only on the mixing angles in matter, and therefore on the densities of the layers. 
Thus, a parametric resonance arises whenever the oscillation phases in both layers produce a constructive interference that leads to the complete conversion.
The solutions can only be fulfilled when
\begin{align}\label{eq:ConPR}
	\cos 2\theta_{\rm C} \leq 0 \text{~and~} \cos 2(2\theta_{\rm M}-\theta_{\rm C}) \geq 0\quad\text{or}\quad
	\cos 2\theta_{\rm C} \geq 0 \text{~and~} \cos 2(2\theta_{\rm M}-\theta_{\rm C}) \leq 0.
\end{align}
If the neutrino energy is above the MSW resonance in the second layer, these translate to  $\theta_{\rm C}\leq 2\theta_{\rm M}\leq \pi/4+\theta_{\rm C}$.
Conversely, below the MSW energy one would require $\pi/4+\theta_{\rm C}\leq 2\theta_{\rm M}\leq \pi/2+\theta_{\rm C}$. 

\begin{figure}[t]
\begin{center}
\includegraphics[width=1\linewidth]{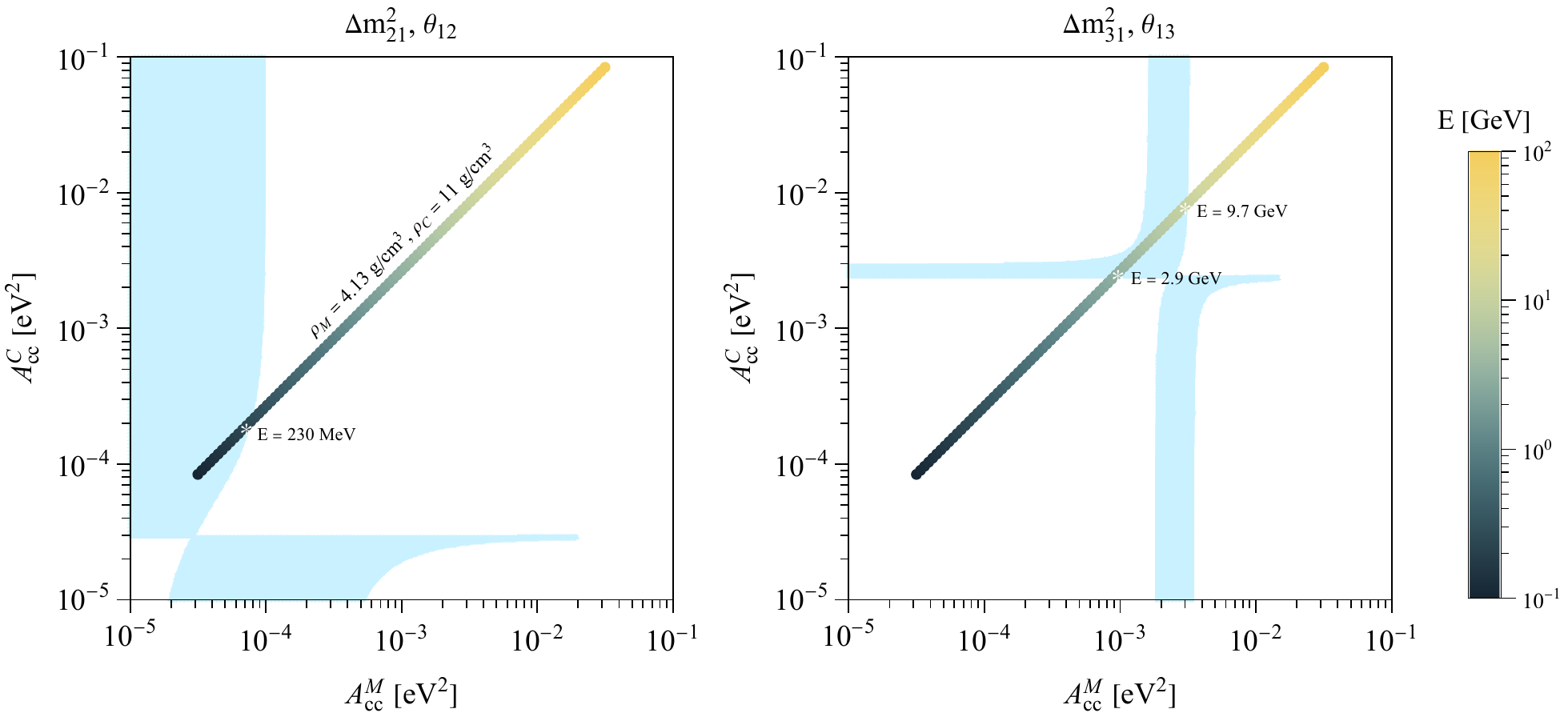}
\caption{Values of $A_{\rm CC}^X$, $X=\mathrm{M,C}$, that could lead to a parametric resonance assuming the solar (left) and atmospheric (right) oscillation parameters. The light blue region corresponds to the values that fulfill the conditions~\eqref{eq:ConPR}. The diagonal colored region corresponds to the values of $A_{\rm CC}$ that are present for the core and mantle densities ($\rho_{\rm C}=11~{\rm g \,cm^{-3}},\ \rho_{\rm M}=4.13~{\rm g \,cm^{-3}}$)  as function of the neutrino energy. \label{fig:ConPR}} 
\end{center}
\end{figure}

We present in Fig.~\ref{fig:ConPR} the regions in light-blue where the necessary conditions \eqref{eq:ConPR} are satisfied in the plane $A_{\rm CC}^{\rm M}$ vs $A_{\rm CC}^{\rm C}$, where $A^X_{\rm CC}$ is the matter contribution in each layer, assuming the two distinct sets of oscillation parameters: solar parameters $\Delta m_{21}^2$ and $\theta_{12}$ in the left; and atmospheric parameters $\Delta m_{31}^2$ and $\theta_{13}$ in the right\footnote{For definiteness, we are assuming normal mass ordering, that is, $\Delta m^2_{31}>0$.}. 
We also color code the values of the $A_{\rm CC}^X$ as a function of the neutrino energy and fixing the densities of core and mantle to $\rho_{\rm C}=11$~g/cm$^3$ and $\rho_{\rm M}=4.13$~g/cm$^3$.
The latter is the average density between the upper and lower mantle. 
We observe that the conditions can be satisfied for the solar parameters in the low energy regime, $E\lesssim 230~\MeV$; meanwhile for the atmospheric sector, the energies are more restricted, $2.9~\GeV\lesssim E \lesssim 9.7~\GeV$.
The important message here is that the regions where parametric resonances take place depend on the matter densities of both layers, and thus the observation of large flavor conversions, or the lack of it, provides an invaluable tool for the determination of such densities with atmospheric neutrino oscillations.

In the region where the parametric resonance can take place, the neutrino baselines required for a full conversion can be estimated from the solutions in Eq.~\eqref{eq:SolPR}. 
Let us note, however, that for atmospheric neutrinos the traveled distances in each layer are not independent, but instead they depend on the zenith angle at which neutrinos cross the Earth. 
For a two-layered model of the Earth, where the mantle layers are combined into one, we have that the baselines on the mantle $L_{\rm M}$ and core $L_{\rm C}$ are related to the zenith angle $\zeta$ as
\begin{align}
 L_{\rm M} = R_{\scriptscriptstyle\oplus}\left(-\cos\zeta-\sqrt{f_c^2-\sin^2\zeta}\right),\qquad L_2 = 2R_{\scriptscriptstyle\oplus}\sqrt{f_c^2-\sin^2\zeta},
\end{align}
where $f_c\equiv R_{\rm core}/R_{\scriptscriptstyle\oplus}$, $R_{\rm core}$ being the core radius. 
Thus, for a given set of densities, we can determine the values of neutrino energies and zenith angle at which a parametric resonance occurs.
Such values are by no means unique; there are several parameters which could produce such resonances. 

\begin{figure}[t]
\begin{center}
\includegraphics[width=0.49\linewidth]{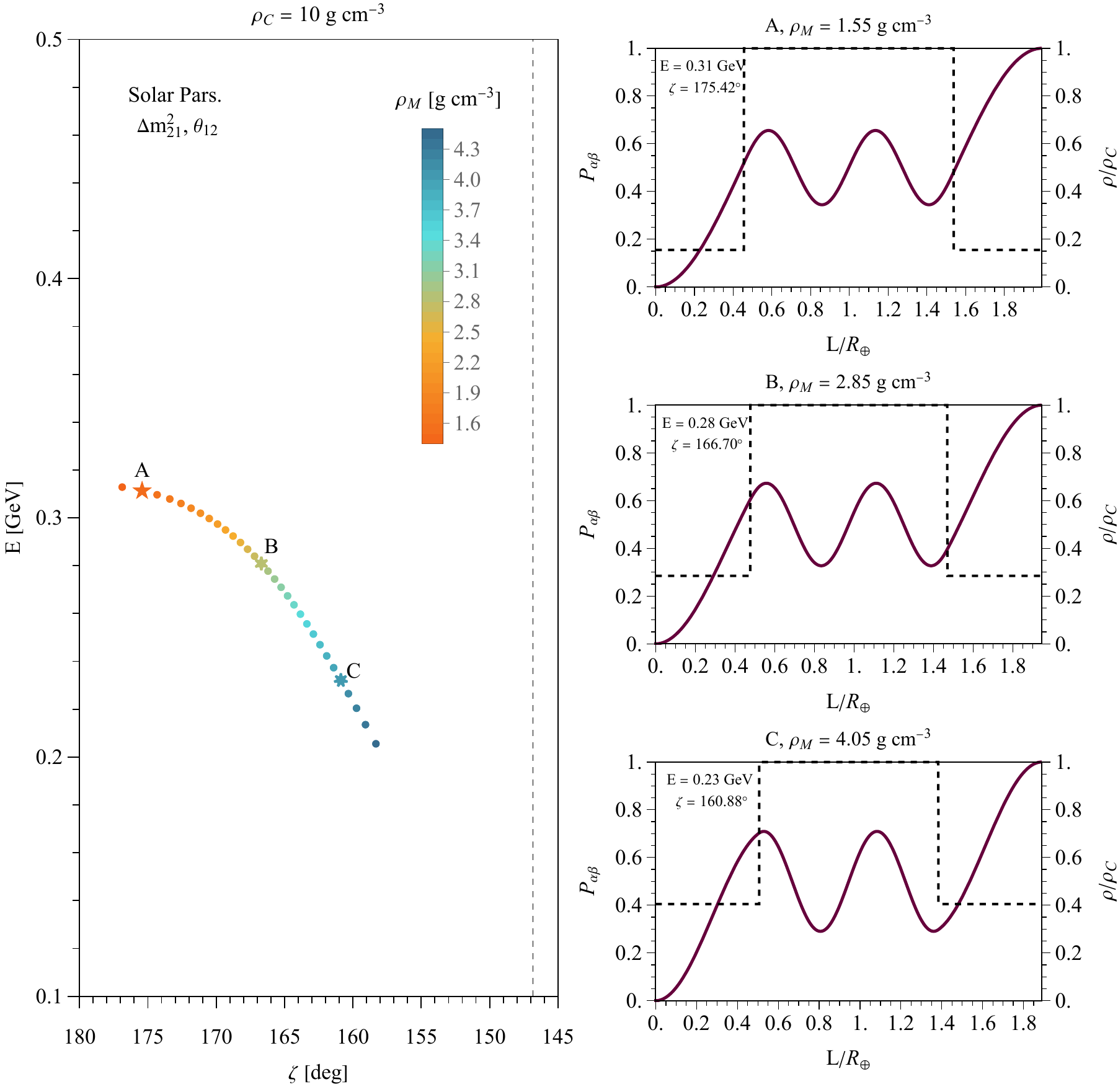}
\includegraphics[width=0.49\linewidth]{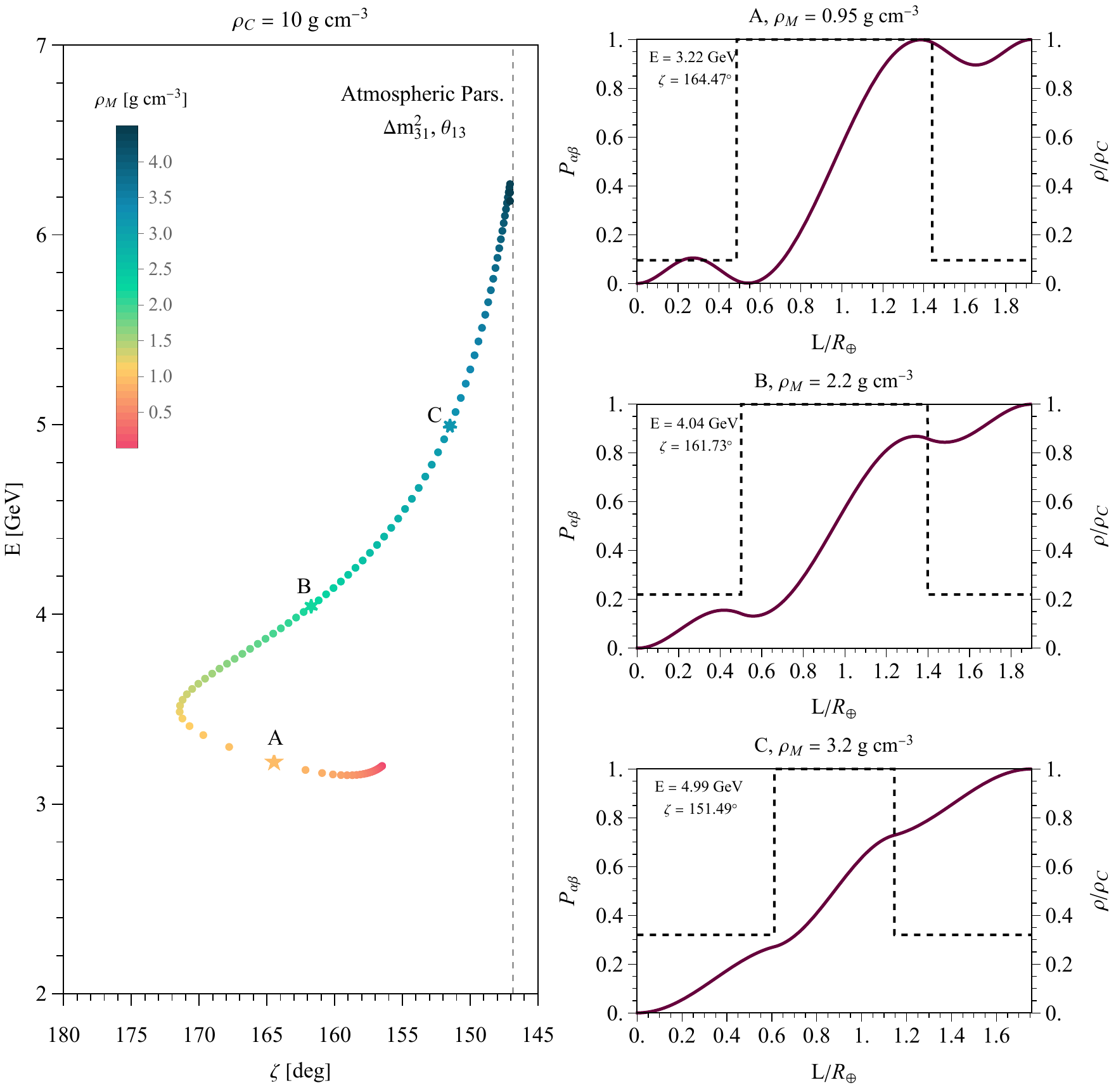}
\caption{Set of neutrino energies and zenith angles that produce a parametric resonance for several values of the mantle density for the solar (left) and atmospheric (right) driven oscillations. In both cases we fix the core density to be $\rho_{\rm C}=10~{\rm g\ cm^{-3}}$. We also present three specific cases as ``A'' (five-pointed star), ``B'' (six-pointed star), and ``C'' (seven-pointed star), together with the probablity $P_{\alpha\beta}$ as function of the travelled distance inside the Earth. The dashed lines indicate the matter profile normalized w.r.t.~the core density. \label{fig:DepRhoPR}}
\end{center}
\end{figure}
To appreciate the dependence of the parametric resonance on the matter densities of the distinct layers, in Fig.~\ref{fig:DepRhoPR} we show a family of solutions as a function of the core matter density that would lead to a full parametric conversion ($P_{\alpha\beta}=1$) for given neutrino energies and zenith angles  for the solar (large left panel) and atmospheric (large right panel) sectors. 
In these, we have  kept the core density fixed at 10~g$/$cm$^3$. 
As we can see, the zenith-energy location of the parametric conversion presents a strong dependence with the core density.

For oscillations driven by solar parameters, we observe that for larger values of the mantle density, we require smaller values of the neutrino energy and larger values of the zenith angle. 
This is required because by increasing the mantle density, the mass splitting in matter increases, while the amplitude decreases --- these energies are above the MSW resonance in both mantle and core --- so  $L_{\rm M}/E$ has to increase to maintain the phase relations that produce the parametric resonance.
To show how the parametric conversion takes place, we show three representative oscillation curves (smaller panels) for each solar (left) and atmospheric (right) sectors, as labeled in the large panels.
The appearance probabilities are given as function of the traveled distance $L$ inside the Earth.
Indeed, the three examples, ``A'', ``B'', and ``C'', have a similar behavior: the phase in the mantle is less than $\pi/2$ while in the core is $3\pi/2\lesssim \phi_{\rm C}\lesssim 2\pi$, so that the core \emph{boosts} the oscillation leading to a full conversion.

For the atmospheric sector, the behavior is more evolved. 
For small densities in the mantle ($\rho_{\rm M}\lesssim 1~{\rm g\ cm^{-3}}$), the energies that lead to a resonance are $E\lesssim 3.5~\GeV$. 
A MSW resonance occurs in the core  for such energies, so the phase required in the mantle is close to $\pi$ in order to preserve the MSW effect, see the  example ``A'' in right side of Fig.~\ref{fig:DepRhoPR}. 
In fact, the phase in the core is $\sim \pi/2$, thus leading to a maximum conversion.
When the matter density in the mantle increases, so that the matter effects are much more relevant, the phase becomes smaller than $\pi$, and the resonance appears because of the interplay between the mantle and core (example ``B'').
Interestingly, the baseline in the core required for the parametric enhancement becomes smaller for higher energies. 
In the extreme cases ($\rho_{\rm M}\gtrsim 3.2~{\rm g\ cm^{-3}}$), the oscillation becomes dominated by matter effects in the mantle, so that the required energies are close to the MSW values.
Let us restate that in the figure above we only show one family of solutions for the parametric resonance condition. 
There are other values of energies and zenith angles that lead to  parametric enhancements besides those presented above.

Although the solutions in Eq.~\eqref{eq:SolPR} are general and lead to a full conversion, there are other combinations of phases and angles that produce \emph{local} maxima~\cite{Akhmedov:1998ui,Akhmedov:1998xq,Chizhov:1998ug,Chizhov:1999az,Chizhov:1999he} as opposed to full flavor conversion. 
Such maxima could be present even in the cases where the conditions in Eq.~\eqref{eq:ConPR} for a full conversion are not fulfilled or the phases $\phi_X$ that come from the solutions are not possible in the system.
In a first type of local maximum, the oscillation phases in \emph{both} layers are integer multiples of $\pi/2$, i.~e.,
\begin{align}\label{eq:FType}
\begin{cases}
	\cos\phi_X= 0, & \text{or}\ \phi_X = \frac{2k + 1}{2}\pi,\quad k = 0,1,2,\ldots, \\
	\cos\phi_Y=0, & \text{or}\ \phi_Y= \frac{2k^\prime + 1}{2}\pi,\quad k^\prime  = 0,1,2,\ldots.
\end{cases}
\end{align}
The resulting probability is very simple and given by
\begin{align}
	P_{\alpha\beta}^{\rm 2f} = \sin^2 2(2\theta_{\rm M}-\theta_{\rm C}).
\end{align}
Note that if indeed $2\theta_{\rm M}-\theta_{\rm C}=(2k+1)\pi/4$, with $k\in \mathbb{Z}$, this leads to full conversion, as it is equivalent to fulfilling the conditions in Eq.~\eqref{eq:ConPR}. 
Nevertheless, even if the oscillation probability is not unity, the flavor conversion still reaches a local maximum.

The second type of local maxima appears when we have a full oscillation cycle in one layer, say $X$, so that the final oscillation probability would come only from the matter effect in the other layer $Y$. This is realized when
\begin{align}
\begin{cases}
	\sin\phi_X= 0, & \text{or}\ \phi_X = k\pi,\quad k = 0,1,2,\ldots, \\
	\cos\phi_Y=0, & \text{or}\ \phi_Y= \frac{2k^\prime + 1}{2}\pi,\quad k^\prime  = 0,1,2,\ldots;
\end{cases}
\end{align}
the probability is then equal to
\begin{align}
	P_{\alpha\beta}^{\rm 2f} = \sin^2 2\theta_Y.
\end{align}
Again, if $\sin^22\theta_Y=1$ we have full flavor conversion, and this would correspond to a MSW resonance in the layer $Y$.
Nevertheless, if the MSW resonance is not achieved, this would still correspond to a local maximum.
The example ``A'' in the right side of Fig.~ \ref{fig:DepRhoPR}, for the atmospheric parameters, is close to this type of maximum, although the phases are not exactly equal to $\pi$ and $\pi/2$ in the mantle and core, respectively.

\vspace{0.5cm}
\textbf{Extension to three-layer Earth:} Now we adjust our analysis to include separate densities in the lower mantle (LM) and upper mantle (UM), which requires us to consider neutrino trajectories that cross three separate layers, as indicated in the right panel of Fig.~\ref{fig:layers}. Here, the evolution operator is
\begin{align}
	{\cal S} &= {\cal S}_{\rm UM}\,{\cal S}_{\rm LM}\,{\cal S}_{\rm C}\,{\cal S}_{\rm LM}\,{\cal S}_{\rm UM} = W_0 \mathbbm{1}_2 - i \vec{\sigma}\cdot\overrightarrow{W},
\end{align}
where ${\cal S}_{\rm UM}$,  ${\cal S}_{\rm LM}$,  ${\cal S}_{\rm C}$ are the evolution operators in the upper mantle, lower mantle and core, respectively. The $W_i$ here are distinct from those in the two-layer case, and can be calculated for the three-layer case. However, we omit them here because their explicit expressions are rather long and not illuminating. As in the two-layer case, a parametric resonance is achieved when $W_0=W_3=0$. Given the intricacy of the expressions, it is not possible to obtain exact solutions to these conditions. Nevertheless, we can solve numerically the system of equations to determine the values of energies and zenith angle required to get a parametric amplification. 

\begin{figure}[t]
\begin{center}
\includegraphics[width=0.485\linewidth]{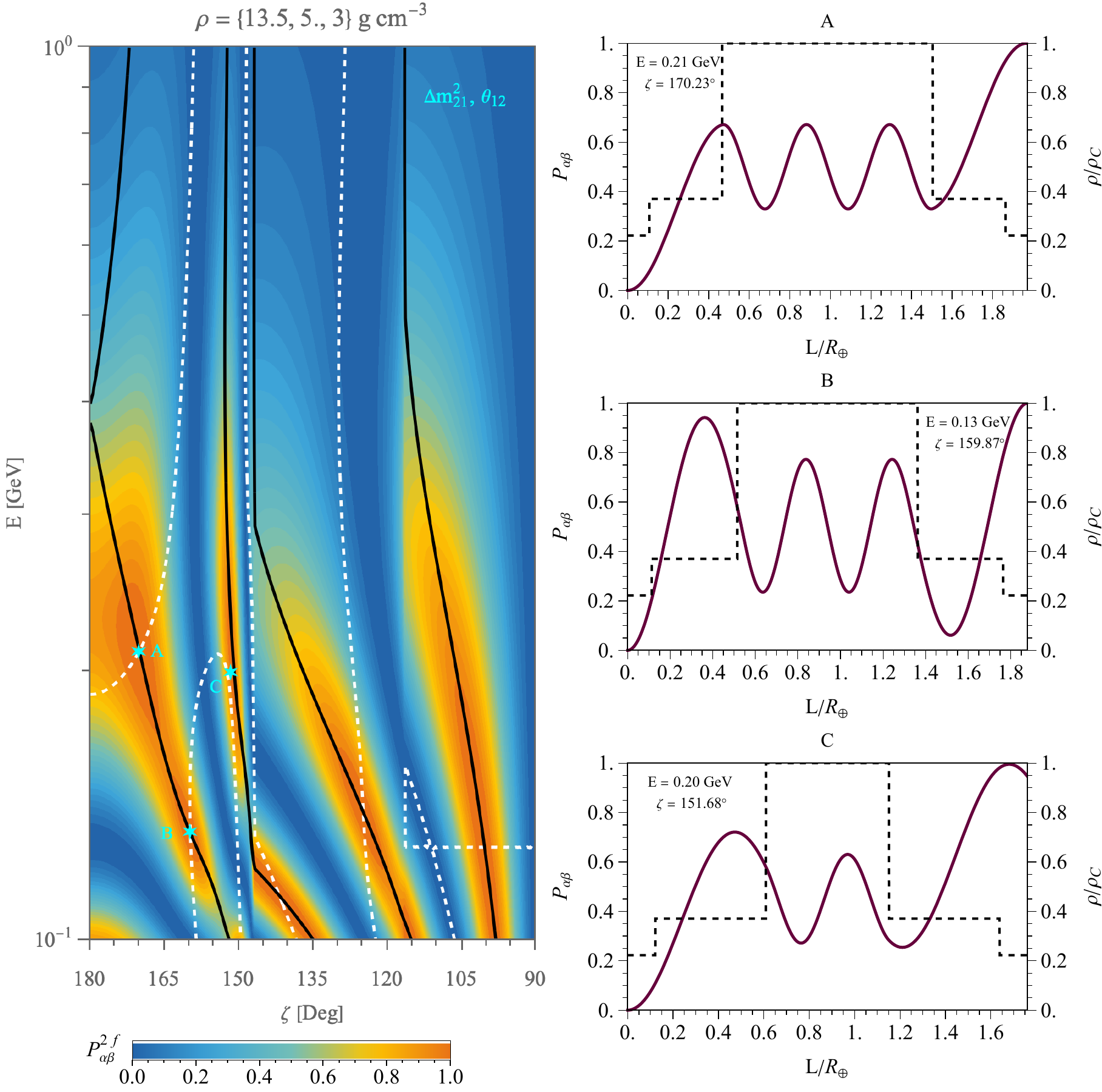}
\includegraphics[width=0.485\linewidth]{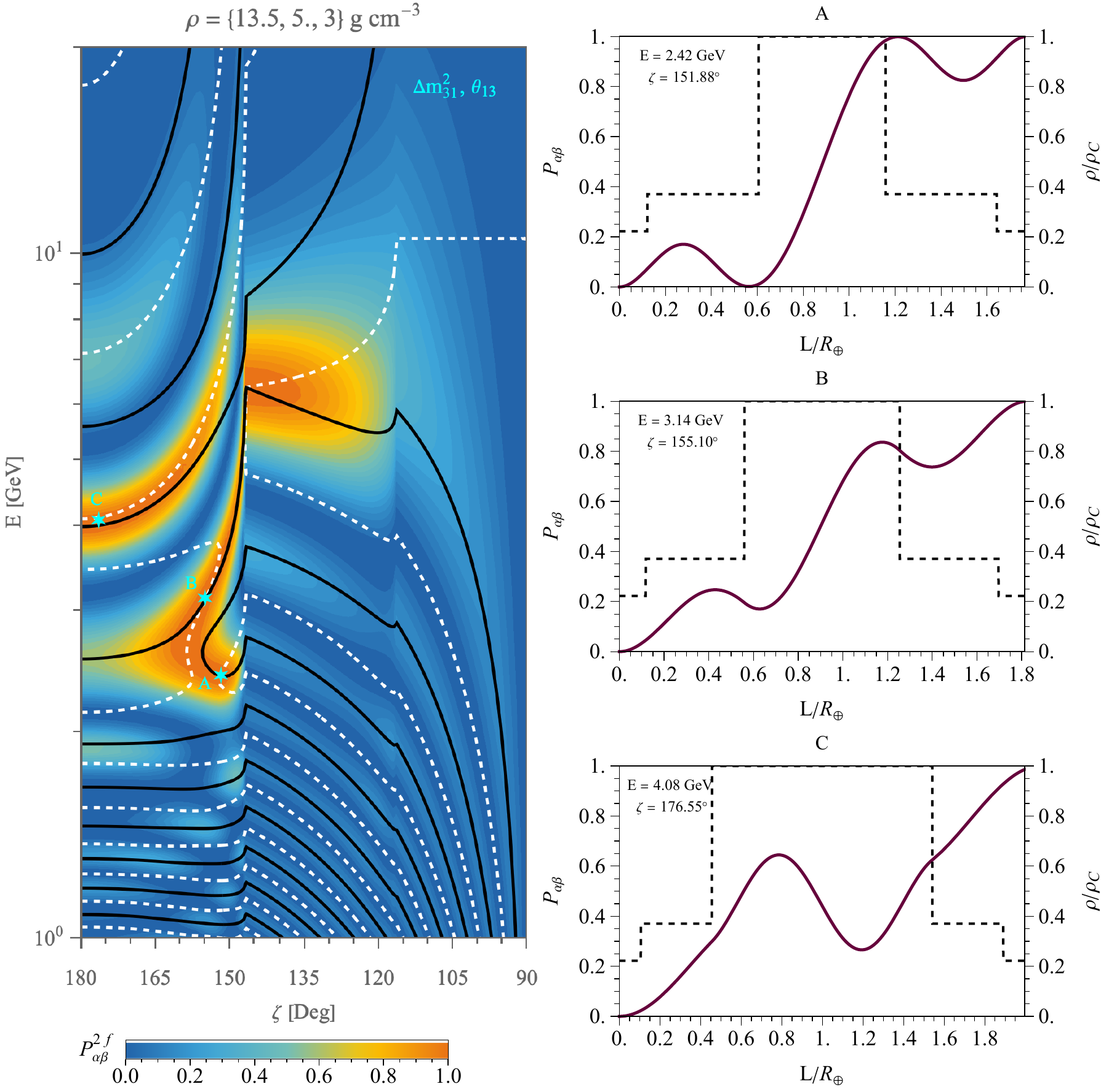}
\caption{Oscillograms for three-layer Earth matter profile with $\rho=\{13.5,\,5,\,3\}~{\rm g\ cm^{-3}}$ for the core, lower mantle, and upper mantle, respectively. The lines represent the parameters where the conditions $W_0=0$ (black), and $W_3=0$ (white dashed) are satisfied. We present three examples where both conditions are (almost) fulfilled: points ``A'' and ``B'', where there is a parametric resonance, and point ``C'' where there is a large amplification in the oscillations. We further present the probabilities as function of the traveled distance inside the Earth together with the matter profile normalized to the Core density. \label{fig:DepRhoPR3l}}
\end{center}
\end{figure}

We present in Fig.~\ref{fig:DepRhoPR3l} an oscillogram for a representative set of densities $\rho=\{13.5,\,5.0,\,3.0\}~{\rm g/cm^{3}}$ for the core, lower mantle, and upper mantle, respectively, for oscillations driven by the solar (left) and atmospheric (right) parameters.
We draw the contours in which $W_0=0$ (black lines) and $W_3=0$ (white dashed lines).
The full parametric conversion happens in the intersection of such contours.
As before, we show three examples of appearance probability as a function of the distance traveled by the neutrino in which  parametric resonances occur.
These correspond to the parameters where the $W_0=0$ and $W_3=0$ contours cross, labeled as ``A'', ``B'', or when they are close to each other (``C'').

Comparing Figs.~\ref{fig:DepRhoPR} and \ref{fig:DepRhoPR3l}, we see that the 2-layer model essentially captures all relevant oscillation physics that takes place in the 3-layer setup.
For the solar sector, in the left panels, the phase developed in the core ranges from $\sim 1.5\pi$ for the point ``C'' to $\sim 2.5\pi$ for the point ``A.''
The sum of phases in case ``A'' in the mantle is about $0.46\pi$, so the resonance appears due to the synergy between both mantle layers and the core. 
In contrast, the total phase in the mantle is larger for points ``B'' and ``C'', $\phi_{\rm UM}+\phi_{\rm LM}\approx 0.7\pi$ due to the larger mantle baseline.
In the core, on the other hand, the phases are reduced relative to ``A'', $\phi_{\rm C}^B\approx 2\pi, \phi_{\rm C}^C\approx 1.32\pi$.
These combinations of phases and mixing angles in matter lead to a parametric resonance in the point ``B,'' while for the point ``C'' there is not a complete conversion, $P_{\alpha\beta} = 0.95$.

For atmospheric parameters, in the right side pf Fig.~\ref{fig:DepRhoPR3l}, we have for point ``A'' that the enhancement comes mainly from the matter effect in the core ---actually an MSW resonance--- since the phase developed in the mantle is $\phi_{\rm UM}+\phi_{\rm LM}\approx\pi$. 
The parametric resonance in point ``B'' relies more in the interplay between mantle and core matter effects, as the phases in both mantles and core are sizable, $\phi_{\rm UM}+\phi_{\rm LM}\approx0.7\pi, \phi_{\rm C}\approx 0.63\pi$, and not a multiples of $\pi$. 
Finally, for point ``C,'' full parametric conversion is almost achieved as the appearance probability is $P_{\alpha\beta} = 0.98$. 
From these numerical studies, we can conclude that the  2-layer approximation works well for both solar and atmospheric sectors.
The reason is the relative smallness of the upper mantle.

\begin{figure}
\begin{center}
\includegraphics[width=0.8\linewidth]{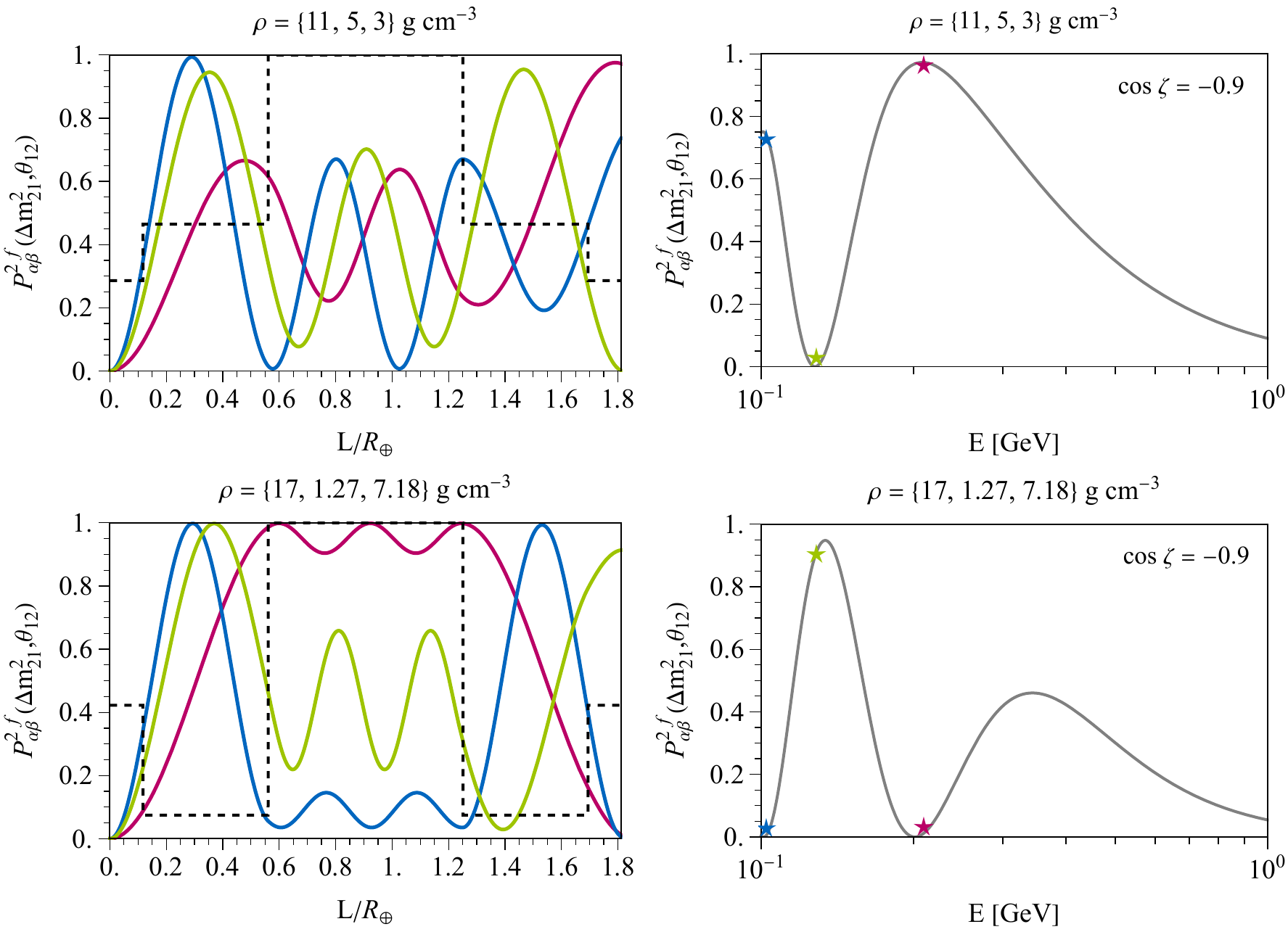}
\caption{Two-flavor oscillation probability $P_{\alpha\beta}^{\rm 2f}$ as function of the travelled distance inside the Earth (left) and energies (right) assuming the true density values of the layers, $\rho=\{11,5.1,3.1\}~{\rm g\ cm^{-3}}$ (top), and values that produce the same Earth mass and moment of inertia $\rho=\{17,1.27,7.18\}~{\rm g\ cm^{-3}}$ (bottom). The oscillation parameters here correspond to the solar values $\Delta m_{21}^2, \theta_{12}$. The red, blue and green lines in the left panels correspond to the stars on the right, and are extrema of the oscillation probability in the true scenario. \label{fig:PSol}}
\end{center}
\end{figure}
\begin{figure}
\begin{center}
\includegraphics[width=0.8\linewidth]{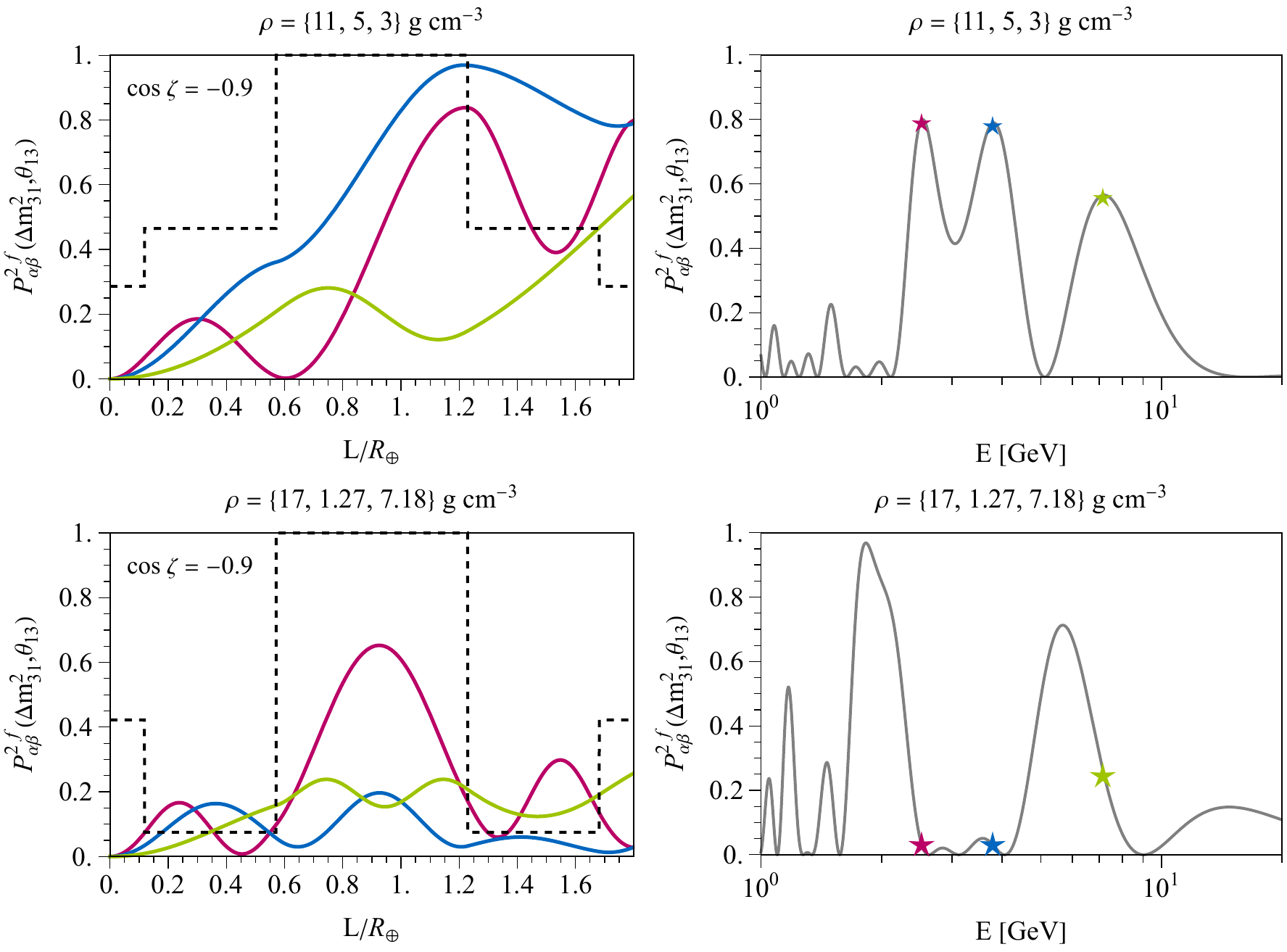}
\caption{Similar to Fig.~\ref{fig:PSol}, but considering atmospheric parameters instead \label{fig:PAtm}}
\end{center}
\end{figure}

To further understand the dependence of the parametric resonance on the densities in this 3-layer case, we compare in Figs.~\ref{fig:PSol} and \ref{fig:PAtm} the oscillation probabilities as a function of the neutrino energy, for solar and atmospheric parameters, respectively, assuming the true values of the densities, $\rho=\{11,5.1,3.1\}~{\rm g\ cm^{-3}}$ (top panels), and values that produce the same Earth mass and moment of inertia $\rho=\{17,1.27,7.18\}~{\rm g\ cm^{-3}}$ (bottom panels).
In all cases we fix the zenith angle at $\cos\zeta=-0.9$. 
We observe that a change in the matter profile can lead to a significant modification on the oscillation pattern. 

For the oscillations generated by solar parameters, see upper right panel in Fig.~\ref{fig:PSol}, we find that for the true density values, there are two local maxima at $E\sim 100~\MeV$ (blue star), $E\sim210~\MeV$ (red star), and a global minimum at $E\sim 130~\MeV$ (green star). 
The origin of such extrema can be inferred from the different values of the phases acquired in the layers, as seen in the upper left panel where the oscillation probability for each point is shown as a function of the baseline (the color of the lines correspond to the color of the stars). 
For the first maximum (blue), the total phase in the mantle is $\sim 0.96\pi$ while in the core is $\sim 1.53\pi$. 
Thus, the final value of the  oscillation comes mostly from the matter effect in the core.

The origin of the second maximum (red) is similar to the one of point ``C'' in the left side of  Fig.~\ref{fig:DepRhoPR3l}. 
The phase in the mantle is $\phi_{\rm UM}+\phi_{\rm LM}\approx 0.57\pi$, while in the core $\phi_{\rm C}\approx 1.37\pi$, so that the enhancement here is purely parametric: the phases and angles conspire to give a close-to-maximal probability.
The absolute minimum (green), on the contrary, appears due to the destructive interference. 

In contrast, the density values in the bottom panel were chosen so that the oscillation probabilities at the same neutrino energies  were interchanged: the maxima become minima, while the minimum is close to the absolute maximum.
This is a result of the modification of the phases and angles in matter, as clearly observed in the left panels. 
The largest modification occurs in the core, where the amplitude of the oscillation is reduced while the phase increases since the density is larger than the true value in this example.

When considering atmospheric parameters in Fig.~\ref{fig:PAtm}, a similar behavior is present. 
For the true density values, top panels, we identify three different maxima (red, blue and green).
Such maxima arise from different phase and angle relations and the interplay between mantle and core densities. 
If we modify the values of the densities in the three layers, again fixing the total Earth mass and moment of inertia, we observe that the oscillation pattern is remarkably altered. 
In fact, due to the modification on the phases two of previous maxima turn into minima, similar to the solar example. 
Maxima appear for other values of the energies, as the parametric conditions would be fulfilled for other parameters given the changes in the densities. 
This serves to show more explicitly that measuring the minima or maxima of atmospheric neutrino oscillations, for both sub-GeV and multi-GeV energies, allows us to constrain the Earth matter profile.

\subsection{Three-flavor neutrino oscillations in the Earth}

Heretofore we have considered a two-flavor system, studying in detail the parametric effects on the oscillations and their dependence on the layer densities. 
The extension to three-flavors has been analyzed in detail in previous literature~\cite{Ohlsson:1999um,Palomares-Ruiz:2004cmm,Akhmedov:2005yj,Akhmedov:2006hb,Akhmedov:2008qt}. 
To avoid repetition, we refer the reader to those works and only quote their take home message.
Given the hierarchy existing between the solar and atmospheric mass splittings $|\Delta m_{31}^2|/\Delta m_{21}^2\approx 34$,  there exist different regions where the oscillation probability is driven by either solar \textit{or} atmospheric parameters. 
Such regions are
\begin{itemize}
	\item \emph{Solar limit.} When the neutrino energies are below that MSW resonance energies for atmospheric parameters, $E\ll E_{\rm MSW}^{\rm atm}$, we can consider the approximation $\sin\theta_{13}\to 0$. 
	In such a case, 
	the appearance probability $P_{\nu_\mu\to\nu_e}$ is
	\begin{align}\label{eq:3fPS}
		P_{\nu_\mu\to\nu_e} 
		\approx\cos^2\theta_{23}P^{\rm 2f}_{\alpha\beta}(\Delta m_{21}^2,\theta_{12}),
	\end{align}
	where $P^{\rm 2f}_{\alpha\beta}(\Delta m_{21}^2,\theta_{12})$ is the two-flavor probability computed in the previous section considering the solar parameters. 
	From this we see that the modification coming from the full three-flavor evolution corresponds to the factor $\cos^2\theta_{23}\approx 0.5$ since $\theta_{23}$ is close to maximal mixing.
	\item \emph{Atmospheric limit.} On the other hand, if we consider the limit when $\Delta m_{21}^2\to 0$ and/or $\theta_{12}\to 0$, 
	the appearance probability simplifies to 
	\begin{align}\label{eq:3fPA}
		P_{\nu_\mu\to\nu_e}
		\approx\sin^2\theta_{23}P^{\rm 2f}_{\alpha\beta}(\Delta m_{31}^2,\theta_{13}).
	\end{align}
	Hence we observe that the correction is similar to the Solar limit, given that $\theta_{23}\approx \pi/4$. 
	Let us notice that this limit is valid when the neutrino energies are above the MSW energy for solar parameters $E_{\rm MSW}^{\rm sol}\ll E$ since $\theta_{12}$ in matter tends to be small. 
	When $E\sim E_{\rm MSW}^{\rm atm}$, the $\theta_{13}$ angle in matter becomes large, and the 1-3 level crossing induces corrections dependent on the solar parameters~\cite{Akhmedov:2008qt}. These corrections are responsible to the slightly mismatch between two and three flavor probabilities in the larger peaks above 1 GeV. 
\end{itemize}

\begin{figure}
\begin{center}
\includegraphics[width=0.9\linewidth]{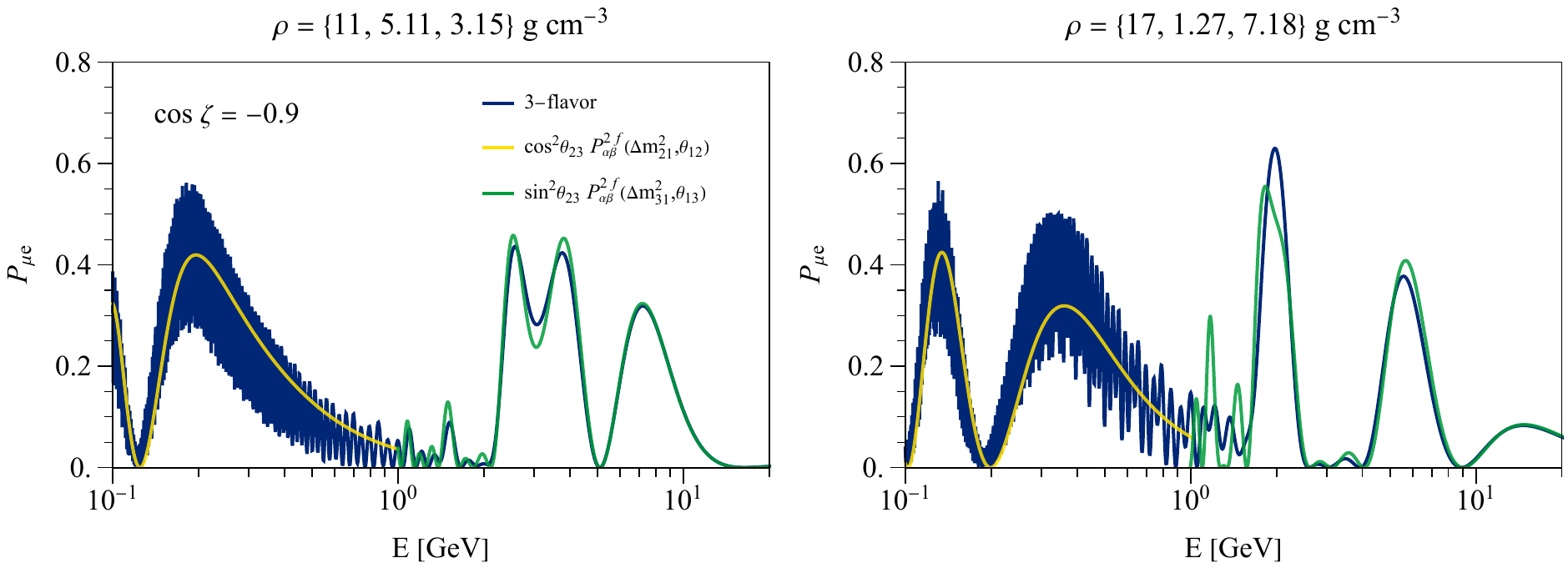}
\caption{Comparison of the full three-flavor appearance oscillation probability $P_{\mu e}$ (blue) with the solar (yellow) and atmospheric (green) approximation from Eqs.~\eqref{eq:3fPS} and~\eqref{eq:3fPA}, respectively. \label{fig:2vs3f}}
\end{center}
\end{figure}
In Fig.~\ref{fig:2vs3f}, we show the approximated probabilities~\eqref{eq:3fPS} (yellow) and~\eqref{eq:3fPA} (green) together with the full three-flavor oscillations (blue) for two sets of densities for the distinct layers of the Earth, true parameters $\rho=\{11,\,5.11,\,3.15\}~{\rm g\ cm^{-3}}$ (left), and the set $\rho=\{17,\, 1.27,\, 7.18\}~{\rm g\ cm^{-3}}$ which leads to the same Earth mass and moment of inertia. 
We have adopted an ad hoc cut on the neutrino energy to separate the two previous regimes at $E=1~\GeV$. 
From the previous figure, we can observe that the approximated probabilities reproduce the main features of the full neutrino oscillation probability, specially the points where the parametric enhancement occurs. 
Specifically, at lower energies, we observe that the maxima and minima from the three-neutrino probability are well described by the two-flavor probability studied before, with the correction coming from the $2-3$ mixing. 
The fast oscillations come from the $\Delta m_{31}^2$ driven oscillations.  

At higher energies, where the solar mixing in matter becomes negligible, we find again that the two-flavor approximation reproduces reasonably well the correct oscillation probability. 
In fact, the energies where the three maxima observed in Fig.~\ref{fig:PAtm} (top panels) are similar to those in the full three-flavor probability (left). The approximation also works for both density sets. 
Thus, we can conclude that the discussion in the previous section describes rather well the dependence on the densities of the three-neutrino oscillations that can be observed in future facilities. 
From now on, we consider the full neutrino oscillations in the three-flavor scenario assuming the three-layer Earth density for the simulations that we will perform.

\begin{figure}[t]
\begin{center}
\includegraphics[width=0.5\linewidth]{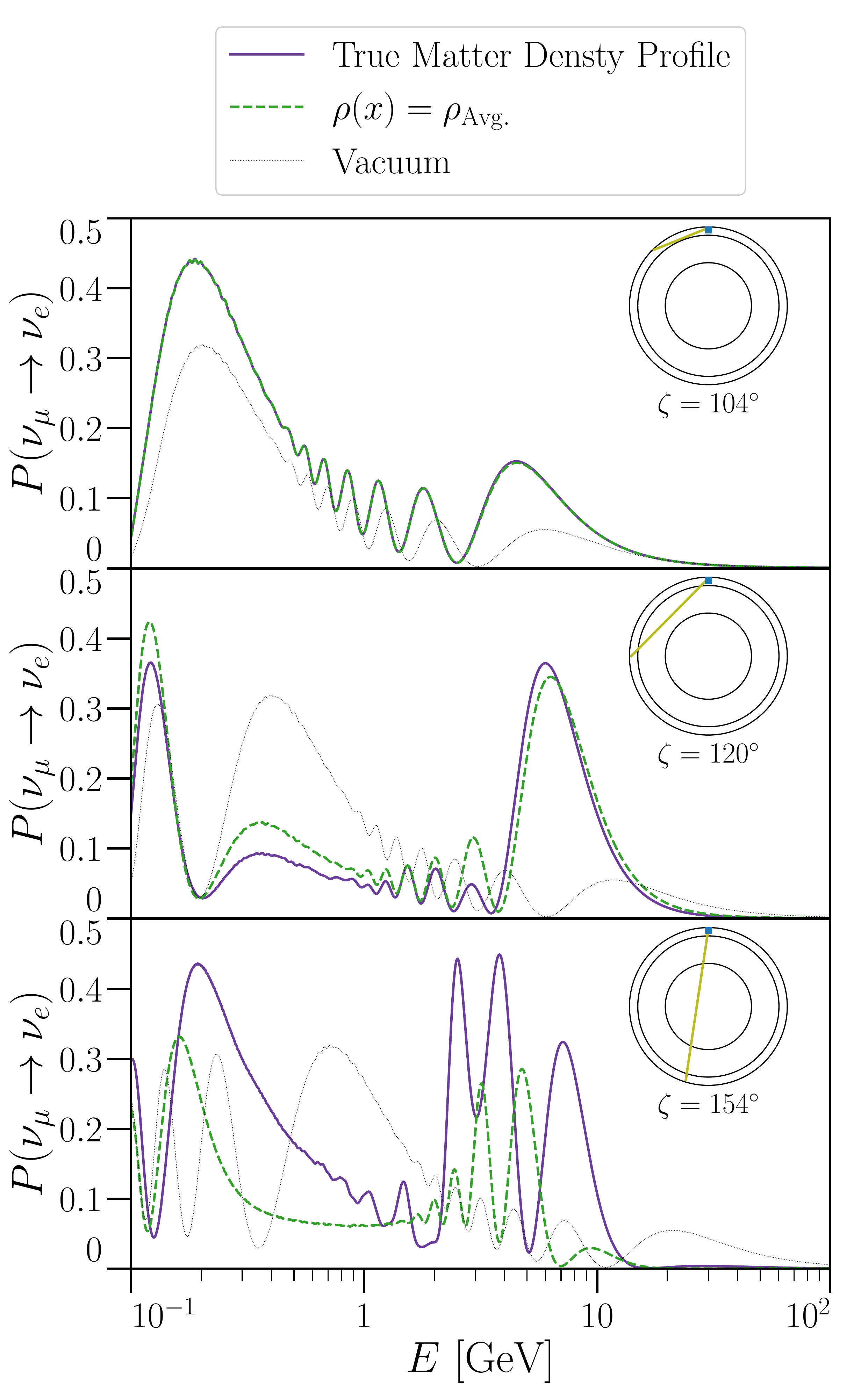}
\caption{Oscillation probability $P(\nu_\mu \to \nu_e)$ for three different zenith angles, $\cos\zeta=-0.25$ (top panel),
$\cos\zeta=-0.5$ (middle panel) $\cos\zeta=-0.9$ (bottom panel) assuming the true mass density profile (MDP) (violet), the average density on the trajectory (dashed green), and vacuum (gray). \label{fig:Pmue_VaryMDP}}
\end{center}
\end{figure}
To firmly establish the importance of the parametric enhancement on atmospheric neutrino oscillations, we present in Fig.~\ref{fig:Pmue_VaryMDP} the full oscillation probabilities for the assumed true mass density profile (MDP) (violet), together with the probability assuming constant matter density equal to the average density on the trajectory (dashed green), and vacuum oscillations (thin gray) for different zenith angles, $\cos\zeta=-0.25$ (top panel), $\cos\zeta=-0.5$ (middle panel) $\cos\zeta=-0.9$ (bottom panel). 
Notice that we have included an energy resolution of 10\% to smear out experimentally unobservable fast wiggles.

For trajectories that only cross the upper mantle, $\cos\zeta=-0.25$, the average density on the trajectory is equal to the density predicted by the MDP since in this case neutrinos only go through one layer. 
When neutrinos transverse the two layers, significant differences appear due to the two-layered density that neutrinos cross. 
For neutrinos crossing the core, we observe immense discrepancies. 
At low energies, the probabilities are enhanced in the region $E\sim 200~\MeV$ compared with the constant density case due the parametric resonance. 
The same parametric enhancement produces maxima in places not expected for a constant density at higher energies. 
The origin of such maxima can be understood in the two-flavor approximation, as seen before. 
Finally, for comparison we observe that neutrino oscillations in vacuum are rather different from both  MPD and the constant density oscillation probabilities. 
The next step to estimate how atmospheric neutrino oscillations can be used to determine the Earth matter profile is to consider the experimental capabilities of DUNE, which will be done in the following section.


\section{Atmospheric Neutrino Reconstruction at DUNE}\label{sec:DUNEDetails}

While not a pillar of the Deep Underground Neutrino Experiment's planned physics programme, DUNE has the potential to observe a plethora of scattering events from atmospheric neutrinos over its ten-plus year lifespan. In comparison to its planned contemporary, Hyper-Kamiokande (HK), this sample will be smaller, due to the relative volumes of the two detectors. 
However, the Liquid Argon Time-Projection-Chamber (LArTPC) technology utilized by DUNE opens up a number of complementary possibilities to the large-statistics capabilities of HK.
Indeed it has been shown that leveraging the LArTPC capabilities of DUNE can significantly enhance several physics searches with atmospheric neutrinos alone, including CP violation measurements~\cite{Kelly:2019itm}, mass ordering determination~\cite{Ternes:2019sak}, and tau neutrino searches~\cite{Conrad:2010mh}.

LArTPCs operate in the following way: when a charged particle travels through argon, it ionizes atoms, freeing electrons that then drift across the uniform electric field established in the detector volume. Those drift electrons are collected on wires at the edge of the detector which use the drift position and timing to reconstruct particle trajectories. As the charged particles travel through the detector, they lose energy from ionization and other effects -- this energy loss is measured as a function of the position along the particle's track and is typically referred to as its $dE/dx$. By identifying and reconstructing the trajectories of particles produced in a neutrino interaction, the LArTPC technology provides a rich, detailed picture of a neutrino scattering event.

This technology exemplifies several key characteristics that aid in studying atmospheric neutrinos, two of which we highlight: the capability to identify the different types of charged particles traveling through the detector volume, and the precise energy/direction measurement of their trajectories, especially those of low-energy particles. Measurements of $dE/dx$ and the topology of energy depositions (track-like vs. shower-like, for instance) are the most important aspects for identifying different particles in the LArTPC. 

Several previous studies, e.g., Refs.~\cite{DeRomeri:2016qwo,Zhu:2018rwc,Friedland:2018vry,Friedland:2020cdp}, have carried out detailed analyses of the DUNE capability of measuring different particles' energies and directions in the LArTPC, both for beam-neutrino-based applications as well as applications for other neutrino sources. In general, charged particles can be reconstructed with $\mathcal{O}$(few-percent) energy resolution and $\mathcal{O}$(few-degree) direction resolution. Notably, in order for a particle to be reconstructed and have its energy/direction measured, it must travel several wire-spacings of distance in the LArTPC volume, translating to a minimum kinetic energy ${\sim}$30 MeV for identification. ArgoNeuT has displayed excellent capabilities, especially in identifying and reconstructing protons, at these low energies~\cite{ArgoNeuT:2018tvi}.

This low-energy capability allows for identification of particles emerging from atmospheric neutrino scattering where the incident neutrino energy is below $1$ GeV. In contrast, HK, which uses Cherenkov light to detect particles, is limited by the Cherenkov threshold (e.g. 1.4 GeV total energy for protons) which hinders analyses at low neutrino energies. Identifying, and properly reconstructing, as many particles as possible in a low-energy atmospheric neutrino interaction allows for the best possible reconstruction of the incoming neutrino energy and direction, critical for the goals of atmospheric neutrino analyses.

Identifying final-state protons in DUNE offers an additional piece of information in these atmospheric analyses -- \textit{statistical} charge identification, enabling \textit{statistical} separation between neutrino and antineutrino samples. When considering charged-current interactions, low-energy neutrinos produce leptons $\ell^-$ and will typically result in a larger number of final-state protons than interactions of antineutrinos that produce $\ell^+$. ArgoNeuT observed this behavior in a beam setting where neutrinos can be focused toward the detector and antineutrinos deflected away, or vice versa. This statistical charge identification allows for us to disentangle effects from, for instance $P(\nu_\mu \to \nu_e)$ and $P(\overline{\nu}_\mu \to \overline{\nu}_e)$ at some level. Further statistical separation is possible by studying the fate of any $\mu^\pm$ created in a charged-current interaction and whether the muons decay into a Michel electron (plus neutrinos) or are captured on an Argon nucleus~\cite{Ternes:2019sak}.

\begin{table}
\begin{center}
\caption{Assumptions of DUNE Far Detector reconstruction and identification capability that enter our analysis. \label{tab:RecoAssumptions}}\vspace{0.1cm}
\begin{tabular}{|c||c|c|c|}\hline
Particle & Minimum K.E. & Angular Uncertainty & Energy Uncertainty \\ \hline\hline
Proton & $30$ MeV & $10^\circ$ & $10\%$ \\ \hline
Pion & $30$ MeV & $10^\circ$ & $10\%$ \\ \hline
$\Lambda$ & $30$ MeV & $10^\circ$ & $10\%$ \\ \hline
$\mu^\pm$ & $5$ MeV & $2^\circ$ & $5\%$ \\ \hline
$e^\pm$ & $10$ MeV & $2^\circ$ & $5\%$ \\ \hline 
\end{tabular}
\end{center}
\end{table}
Altogether, these facets demonstrate how DUNE is capable of amassing a large sample of atmospheric neutrino events, particularly for $E < 1$ GeV, and reconstruct the incoming neutrino direction and energy with some level of precision. In the remainder of this section, we quantify this capability and describe how the reconstruction capabilities enter our simulation and analyses. To perform our simulation, we first quantify the degree to which we can measure individual particles' three-momenta, as well as the minimum kinetic energy for which one could confidently reconstruct and identify a given particle. The assumptions we make, consistent with projections from the DUNE collaboration~\cite{DUNE:2020lwj, DUNE:2020ypp}, are summarized in Table~\ref{tab:RecoAssumptions}.

With these assumptions, we simulate neutrino scattering on Argon using the \texttt{NuWro} neutrino event generator~\cite{Golan:2012rfa}, which accounts for relevant nuclear effects such as Pauli blocking, Fermi motion and two nucleon correlations, as well as hard scattering physics like quasi-elastic transitions, meson exchange currents, resonance production and deep inelastic scattering.
After discarding particles below the thresholds in Table~\ref{tab:RecoAssumptions}, we perform a Gaussian smearing on the particles' directions and energies. We take the visible particles (i.e., discarding neutrons) and sum their four-momenta. The total energy of the visible particles is our proxy for the incoming neutrino energy, $E_{\rm rec}$ and the direction of this sum is our proxy for the incoming neutrino direction, $\zeta_{\rm rec}$. We also divide our samples based on the number of visible protons and pions in the final state, as we will discuss later. Note that we only take into account events with at least one reconstructed muon or electron in the final state, so we are not considering any neutral current events.

\begin{figure}[t!]
\begin{center}
\includegraphics[width=0.65\linewidth]{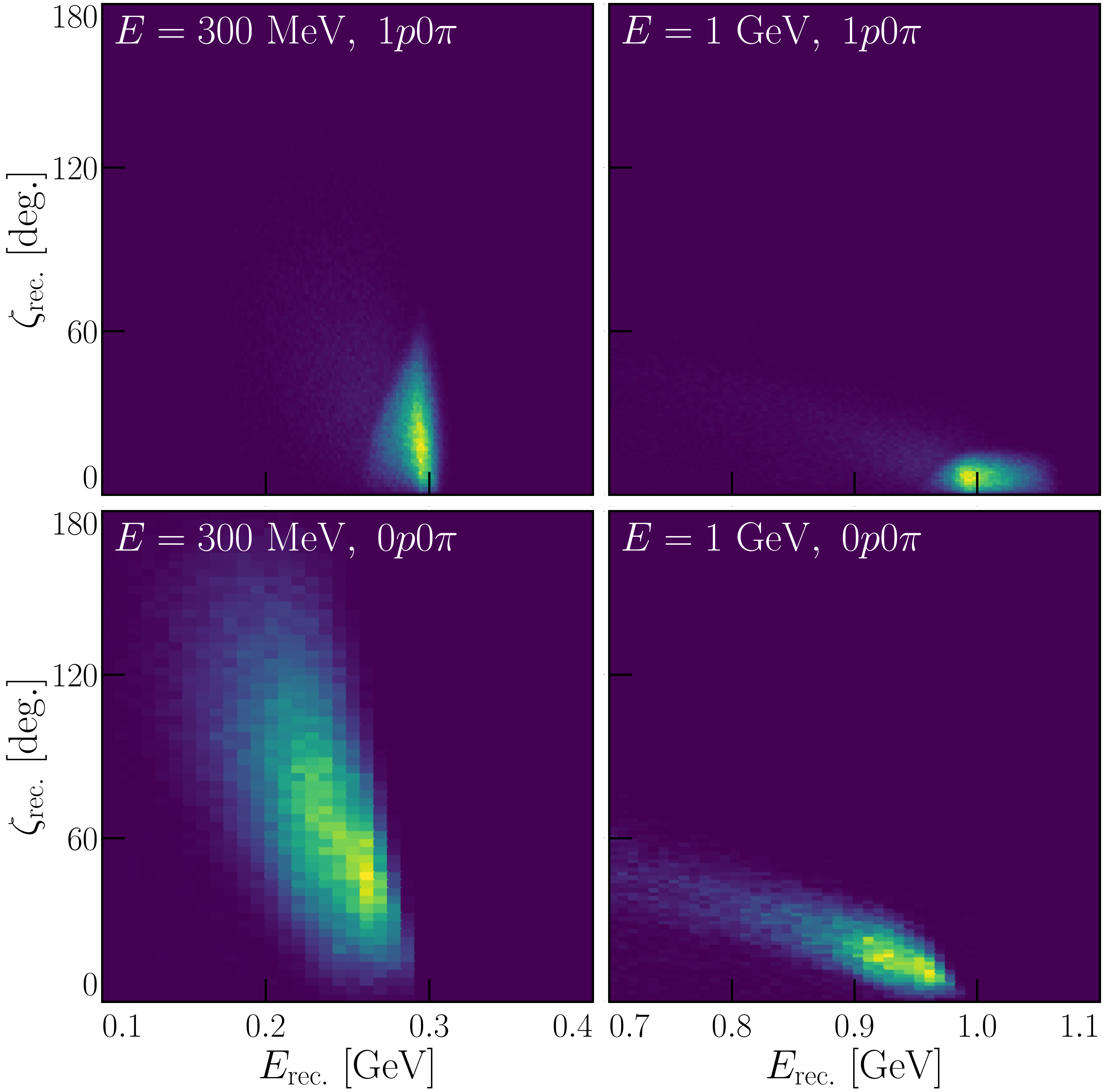}
\caption{Reconstructed energy and direction of neutrinos with 300 MeV (left) or 1 GeV (right) energy, where the top (bottom) row corresponds to final states with one (no) visible proton(s). Yellow regions correspond to where more events will be reconstructed, where purple correspond to fewer events. Note that the angular (y-axis) scale is identical in all four panels, whereas the energy (x-axis) scale changes from the left to right panels. \label{fig:Reco2D}}
\end{center}
\end{figure}
\begin{figure}[b!]
\begin{center}
\includegraphics[width=0.60\linewidth]{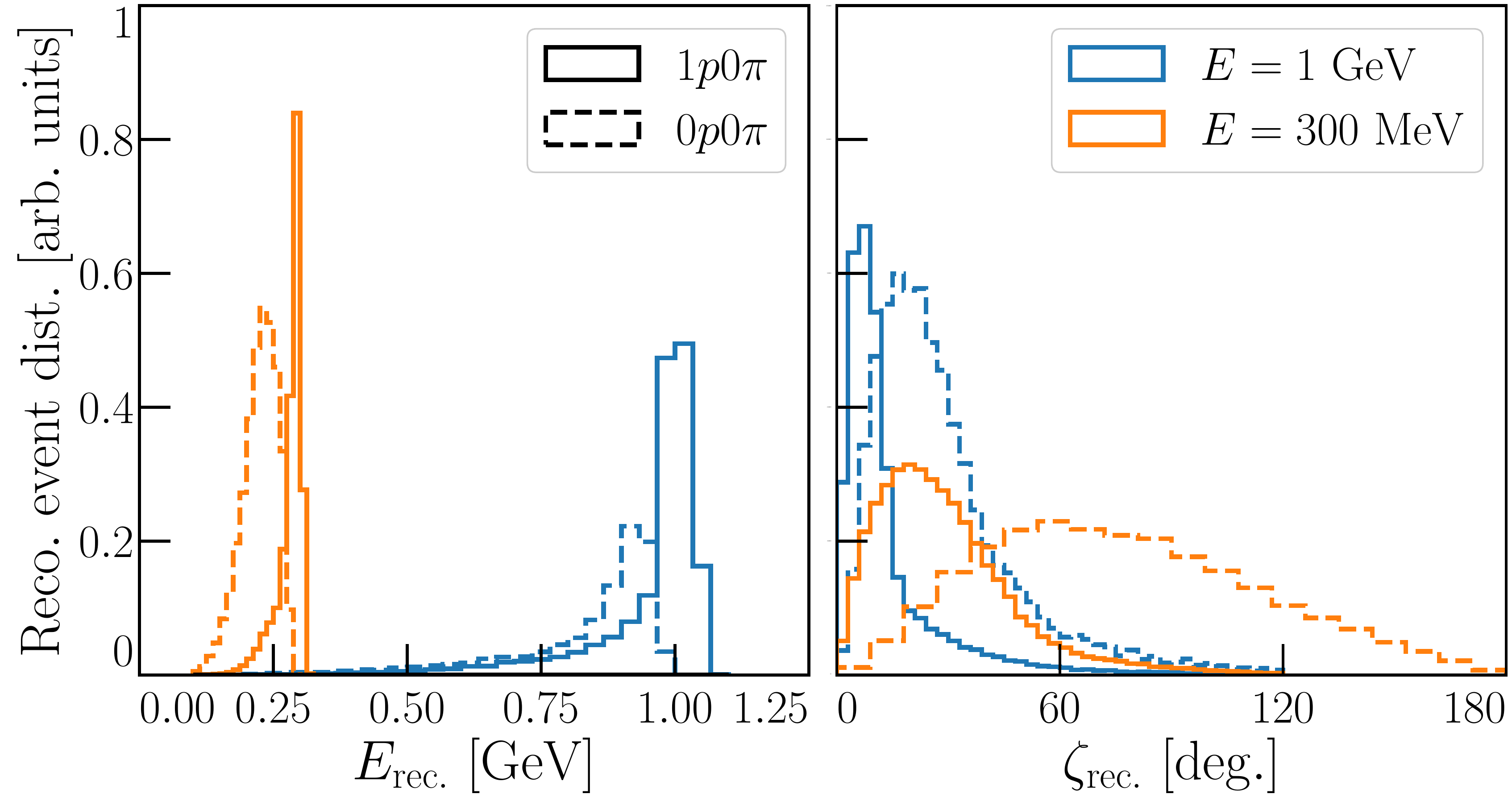}
\caption{One-dimensional distributions of reconstructed energy (left) and direction (right) for neutrinos with $1$ GeV (blue) or $300$ MeV (orange) energy. The solid (dashed) lines are for event samples with one (zero) visible final-state proton(s). See text for more details. \label{fig:Reco1D}}
\end{center}
\end{figure}
Fig.~\ref{fig:Reco2D} presents how well we expect DUNE to reconstruct the incoming energy and direction of atmospheric neutrinos with $E = 300$ MeV (left panels) and $E = 1$ GeV (right panels), with the top  panels corresponding to events with one visible proton in the final state, while the bottom panel has none. The yellow regions correspond to larger  density and therefore more  reconstructed events, while purple corresponds to the opposite.
We highlight several features here -- for all four samples shown, the energy of the incoming neutrino is reconstructed with decent precision, and that precision is better for high-energy neutrinos than lower-energy ones. 
Likewise, reconstructing the incoming neutrino direction is easier for high-energy ones, and having a visible proton in the final state helps greatly in this reconstruction. 
Without a final state proton, there is a correlation between the reconstructed energy and angle.
This bias is present for both lower and higher energy neutrinos.
In our analyses, this reconstruction serves as a migration matrix, mapping $(E_{\rm true}, \zeta_{\rm true})$ onto $(E_{\rm rec}, \zeta_{\rm rec})$. 
Given the energy and angular resolutions, we organize the data in 60 bins of energy, distributed logarithmically from 100~MeV to 100~GeV, and 40 equal angular bins of $4.5^\degree$.

For further comparison, we also take the results of Fig.~\ref{fig:Reco2D} and project them down to either the x- or y-axis and analyze one-dimensional distributions of these quantities in Fig.~\ref{fig:Reco1D}.
We show the expected measurement of the reconstructed neutrino energy (left panel) and of the reconstructed incoming neutrino direction (right panel), for true neutrino energies of 1~GeV (blue) and 300~MeV (orange) and for one-proton (solid) and zero-proton (dashed) final states.
In both panels, the normalization of the distributions is arbitrary and simply for comparison. 
While the information encoded here is the same as in Fig.~\ref{fig:Reco2D}, the angular capability of DUNE is now even more apparent -- the single-proton final state provides angular measurements roughly twice as precise than events with no visible protons. 
If we focus on the 300~MeV case, the angular resolution is larger than 60 degrees, and there is a large bias in reconstructing \emph{the wrong neutrino incoming direction}.
As expected, it is impossible to probe any structure of the Earth's matter profile with this sample, if final state protons are not reconstructed.
The LArTPC capability of identifying protons is therefore critical for the analyses we propose in this work.

\section{Atmospheric Neutrino Fluxes and Uncertainties}\label{sec:FluxUncertainties}

The collision of cosmic rays with atmospheric nuclei generates a
flux of mesons, mainly pions and kaons at lower energies.
The mesons propagate
throughout the atmosphere before arriving at the Earth. 
Along its
propagation, the meson flux will decay into a flux of muons and
neutrinos which can be observed by neutrino experiments in the Earth. 
If the
energies of the muons are low enough ($E_{\mu} \lesssim1$~GeV), they
will also decay before arriving to the Earth and contribute to the
neutrino flux.

\begin{figure}[b]
\begin{center}
\includegraphics[width=1.\linewidth]{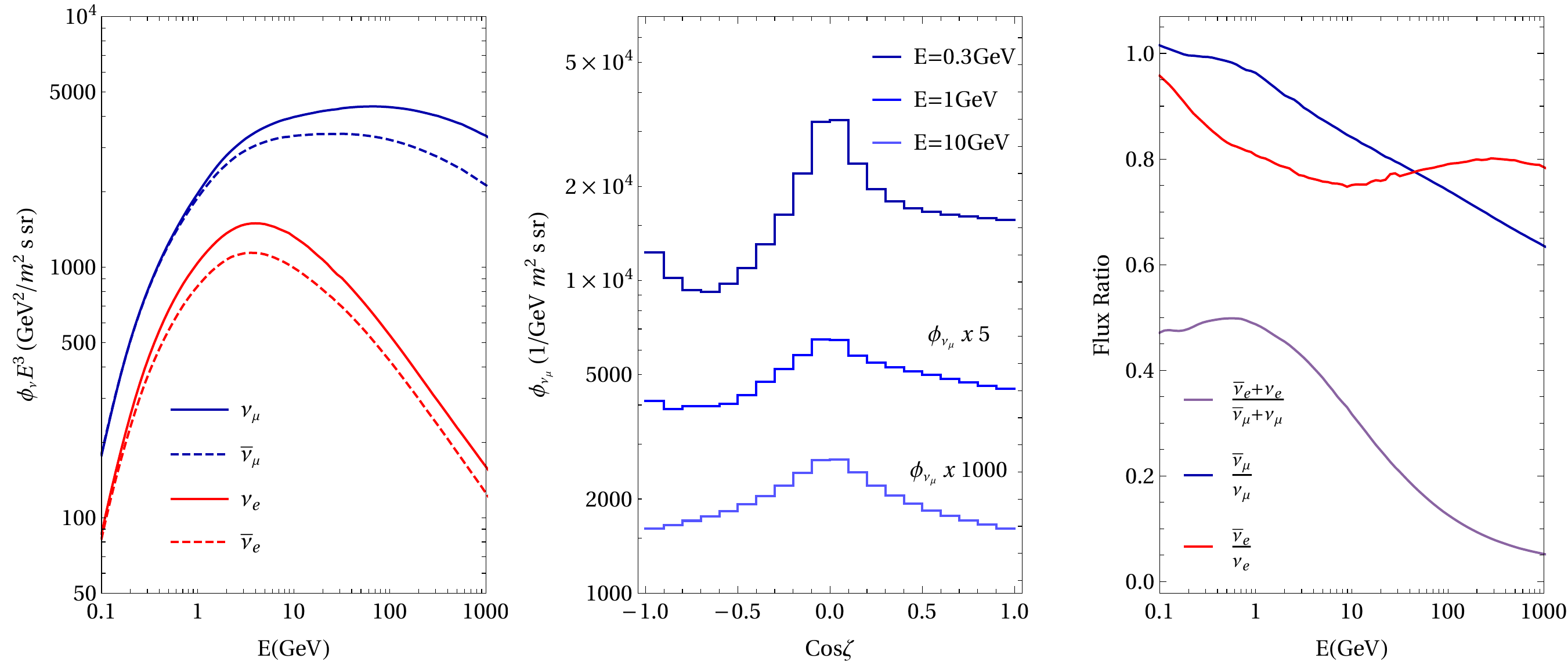}
\caption{\emph{Left}: atmospheric neutrino fluxes as a function of the neutrino energy for $\nu_\mu$ (blue), $\overline\nu_\mu$ (blue dashed), $\nu_e$ (red), $\overline\nu_e$ (red dashed). \emph{Middle}: Zenith dependence of atmospheric neutrino flux for neutrino energies of 0.3~GeV (dark blue), 1~GeV (blue) and 10~GeV (light blue). \emph{Right}: flux ratios for $\nu_\mu+\overline\nu_\mu/\nu_e+\overline\nu_e$ (purple), $\overline\nu_\mu/\nu_\mu$ (blue), and $\overline\nu_e/\nu_e$ (red).\label{fig:Flux}}
\end{center}
\end{figure}
The primary spectrum of cosmic rays arriving at the Earth is composed of free protons ($\sim 80\%$) and bound nuclei ($\sim20\%$)~\cite{Gaisser:2002jj,Dembinski:2017zsh}, and spans many orders of magnitude in energy, from the MeV to the EeV scale.
This atmospheric neutrino flux exhibits an energy dependence of  $\sim E^{-3}$ around $1$~GeV as can be seen in the left panel of Fig~\ref{fig:Flux}. 
In this work, we focus on neutrinos with
energies from $0.1$~GeV to $100$~GeV, which comprises the
most relevant oscillation physics of atmospheric neutrinos.

Although cosmic rays approach the Earth isotropically, the
interaction with solar winds and the Earth's magnetic field induce
several anisotropies in the neutrino flux in the sub-GeV region, see middle panel of 
Fig.~\ref{fig:Flux}. 
The geomagnetic
effect will also modify the propagation of charged mesons generated after the
primary interaction. 
In particular, low-momentum mesons that would fly away from the Earth can be trapped by the magnetic field, which
enhances
the low energy neutrino flux. 
Also, since most cosmic ray particles have positive charge, the geomagnetic effects induce an east-west
asymmetry~\cite{Super-Kamiokande:1999mpf}. 
As the geomagnetic field is location dependent, so it is the atmospheric neutrino flux~\cite{Honda:2015fha}.

An important aspect of atmospheric neutrinos is that their composition changes for different neutrino energy.
At low energies, pions decays to muon neutrinos and muons, which in turn, if below 1~GeV, tend to decay before reaching the surface.
This induces a flavor ratio $\nu_{e} +
\overline{\nu}_{e}/\nu_{\mu} + \overline{\nu}_{\mu} \sim 1/2$, see right panel of Fig.~\ref{fig:Flux}. 
At higher energies, it is more likely that the muon reaches the surface before decaying, which suppresses this flavor ratio.
Additionally, there is an asymmetry between the neutrino and antineutrino fluxes, due to typical positive charges of cosmic rays.
Nevertheless, at low energies the geomagnetic field traps secondary particles, leading to more collisions in the atmosphere, raising the particle multiplicity and washing out this asymmetry.
As the neutrino
energy increases, the primary mesons created by the proton interaction
dominates the contribution to the neutrino flux and
$\overline{\nu}/\nu$ decreases. 

Moreover, the thickness of the atmosphere along the meson/muon trajectory and the geomagnetic field induce an asymmetry on the zenith distribution of the neutrino flux. 
This effect can be seen in the middle panel of Fig.~\ref{fig:Flux} where the muon neutrino flux is shown as a function of the zenith angle for different neutrino energies.

In this work, we use the tables from Ref.~\cite{Honda:2015fha} that are
based on a 3-dimensional simulation of the atmospheric neutrino flux.\footnote{We use the tables corresponding to the atmospheric neutrino flux at the location of Super-Kamiokande, which shares a similar latitude to the future DUNE experiment. While the different effects (magnetic fields, etc.) could yield a slightly different atmospheric neutrino flux in South Dakota, no publicly available version of this exists. We encourage the DUNE collaboration to include detailed simulations of the atmospheric neutrino flux in their future studies.}
The simulation includes the transverse momentum for
all secondary particles as well as geomagnetic field effects. 
The 3-dimensional
propagation of the neutrino flux is more relevant in the sub-GeV
region where there is less correlation between the meson and cosmic ray directions. 

As can be suspected from the discussion above, the determination of the atmospheric neutrino flux is far from trivial and plagued by systematic uncertainties.
The main uncertainties are related to the cosmic ray flux and the hadron production, while others, such as the density of the atmosphere and those associated to the geomagnetic field, are relatively small~\cite{Barr:2006it}. 
To account for such uncertainties in a realistic way, we start by parameterizing the
atmospheric neutrino flux for the flavor $\alpha$ with enough freedom to implement energy and zenith uncertainties, namely,
\begin{equation}
  \Phi_{\alpha}(E,\cos\zeta) =  f_{\alpha}(E,\cos\zeta)\Phi_{0}\left(\frac{E}{E_{0}}\right)^{\delta} \eta(\cos\zeta),
\end{equation}
where $f_{\alpha}(E,\cos\zeta)$ is the flux of atmospheric neutrinos of flavor is a function of both energy and zenith angle~\cite{Honda:2015fha}.
The remaining factors play the following role:
$\Phi_{0}$ describes the unknown  normalization of the flux; $(E/E_{0})^{\delta}$ induces a spectral tilt, accounting for energy dependence uncertainties; and $\eta(\cos\zeta)$ describes the zenith distribution uncertainty.
In particular, the $\eta(\cos\zeta)$ is chosen to be
\begin{equation}\label{eq:zenith_uncertainty}
  \eta(\cos\zeta) \equiv \left[1 - C_{u}\tanh(\cos\zeta)^{2}\right]\Theta(\cos\zeta)
  				+\left[1 - C_{d}\tanh(\cos\zeta)^{2})\right]\Theta(-\cos\zeta),
\end{equation}
where $\Theta$ is a Heaviside step function.

Let us discuss the dominant systematic uncertainties in the atmospheric neutrino fluxes. 
Following Ref.~\cite{Barr:2006it}, 
the normalization uncertainties of the flux itself, due to uncertainties on the cosmic-ray spectrum, range from 10\% to 40\% in the $0.1$~GeV to the 100~TeV region.
Ratios of flavors on the other hand tend to be better known.
At lower energies, $\overline{\nu}_{e}/\nu_{e}$ exhibits the
largest uncertainty of about $6\%$ since each component is produced by a
different meson decay chain. 
The $\overline{\nu}_{\mu}/\nu_{\mu}$ and
$\nu_{e}/\nu_{\mu}$, ratios are better estimated as these flavors are mostly sourced by the pion decay chain. 
At higher energies, the interaction of
the up-going and down-going muons with the Earth will increase the
uncertainty of $\overline{\nu}/\nu$ along those directions to about $20\%$, and the uncertainty of $\nu_{e}/\nu_{\mu}$ to $\sim5\%$ for
neutrino energies near 100~GeV. 
In our simulations, to be on the conservative side, we adopted the following uncertainties for those quantities for the full energy range: 
overall normalization of the flux of $40\%$; 
$\nu_{e} + \overline{\nu}_{e}$ and $\nu_{\mu} +
\overline{\nu}_{\mu}$ flux ratios of $5\%$; 
neutrino to antineutrino ratio of $2\%$; 
an absolute uncertainty of $0.2$ in the
energy distortion parameter $\delta$; and a zenith distortion parametrized by $C_{u,d}=0\pm0.2$. The uncertainties used in the analysis are summarized in Table~\ref{tab:Fluxuncert}.

\begin{table}
\begin{center}
\caption{Uncertainties and priors in the atmospheric neutrino flux used in our analysis. \label{tab:Fluxuncert}}\vspace{0.1cm}
\begin{tabular}{|c|c|}\hline
Systematic  &Uncertainties/Priors \\ \hline\hline
Normalization ($\Phi_{0}$) & $40\%$ \\ \hline
Flavor ratio ($\nu_{e}/\nu_{\mu}$) & $5\%$ \\ \hline
Neutrino to antineutrino ratio ($\overline{\nu}/\nu$)  & $2\%$ \\ \hline
Energy distortion ($\delta$)   &$0\pm0.2$ \\ \hline
Zenith distortion ($C_{u,d}$)   &$0\pm0.2$ \\ \hline
\end{tabular}
\end{center}
\end{table}

As a final comment, our simulation shows that all parameters related to the systematic uncertainties can be measured at DUNE with a precision which is better than what we adopted as priors. 
This is important because otherwise one could suspect that the experimental sensitivity to a given physics measurement could be  driven by the chosen prior on the uncertainties (see also Ref.~\cite{Kelly:2019itm}).
We now proceed to the results of our simulation on DUNE's sensitivity to measure the Earth's matter profile.

\section{Results and Discussion}\label{sec:ResultsDiscussion}
We now proceed to the numerical results of our analysis.
First, we will estimate how well DUNE can measure the total mass of the Earth, keeping the shape of the matter profile fixed.
Then we will estimate DUNE's sensitivity to the matter profile itself accounting for current measurements of the Earth's total mass and moment of inertia.
Finally we will repeat the second analysis but discarding the total mass and moment of inertia measurements.

\subsection{Overall Earth Mass Measurement}
\label{subsec:MeasureMass}
Let us start with the simplest measurement of Earth's properties that can be performed with atmospheric neutrinos at DUNE: the determination of the total mass of the Earth.
To simplify the calculation, we consider the 3-layer Earth model introduced in Section~\ref{sec:AtmoNu}.
Although there are detailed models of the matter density profile available, such as the Preliminary Reference Earth Model (PREM)~\cite{Dziewonski:1981xy}, neutrino oscillation data will not be able to distinguish fine-grained matter density models.
This measurement will allow us to assess DUNE's sensitivity with a simple figure of merit.

Our statistical treatment follows the $\chi^2$ defined in Ref.~\cite{Kelly:2019itm}, with the addition of the zenith-dependent uncertainty defined in Eq.~\eqref{eq:zenith_uncertainty}.
Concomitantly with atmospheric neutrino data taking, the beam neutrino program in DUNE will be providing the most precise measurements of the mixing angle $\theta_{23}$, the $CP$ violating phase and the neutrino mass ordering.
Since all oscillation parameters will be better measured with DUNE's beam neutrinos or other neutrino experiments, such as the future Hyper-Kamiokande~\cite{Hyper-Kamiokande:2018ofw} and JUNO~\cite{JUNO:2015sjr} experiments; or the current Daya Bay~\cite{DayaBay:2018yms}, KamLAND~\cite{KamLAND:2013rgu}, and solar neutrino experiments~\cite{Aharmim:2011vm, Borexino:2013zhu, Super-Kamiokande:2016yck}, we treat those parameters as constant.
We have chosen the global best fits  to neutrino oscillation data of Ref.~\cite{Esteban:2020cvm} for either normal ordering, that is, $\Delta m^2_{21}=7.42\times10^{-5}$~eV$^2$, $\sin^2\theta_{12}=0.304$, $\Delta m^2_{31}=2.51\times10^{-3}$~eV$^2$, $\sin^2\theta_{13}=2.22 \times 10^{-2}$, $\sin^2\theta_{23}=0.570$ and $\delta_{CP}=195^\degree$; or inverted ordering, $\Delta m^2_{21}=7.42\times10^{-5}$~eV$^2$, $\sin^2\theta_{12}=0.304$, $\Delta m^2_{32}=-2.50\times10^{-3}$~eV$^2$, $\sin^2\theta_{13}=2.24 \times 10^{-2}$, $\sin^2\theta_{23}=0.575$ and $\delta_{CP}=286^\degree$.

There is one further detail worth mentioning regarding DUNE's measurement.
Neutrino oscillation is affected not by the Earth's density itself, but by its electron number density, and thus the measurement really provides the total electron number of the Earth.
In relating the electron number density to the mass density, we have assumed the proton-to-neutron fraction to be 1 (which is close to that estimated in Ref.~\cite{Dziewonski:1981xy}), along with the fact that the Earth is essentially electrically neutral.
A mismatch between the total mass of the Earth measured with neutrino oscillations and other techniques could, in principle, indicate a problem with the assumed Earth's chemical composition.
Keeping that in mind, we will hereafter discuss the matter density measurements with the implicit proton-to-neutron ratio assumption.

\begin{figure}
\begin{center}
\includegraphics[width=0.8\linewidth]{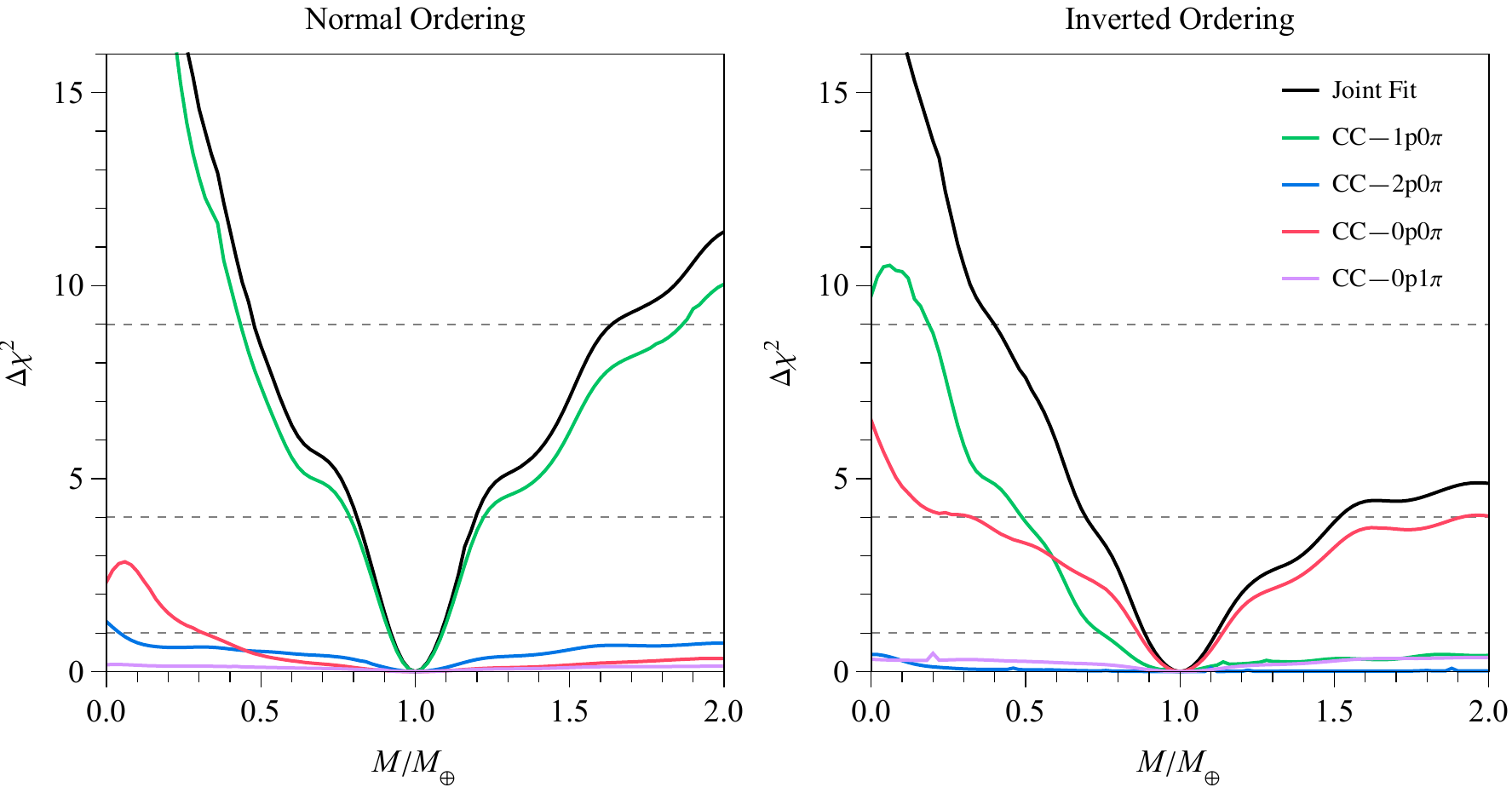}
\caption{DUNE's sensitivity to the total mass of the Earth for several different charged current (CC) event topologies accounting for different number of protons and pions (colored lines as indicated) and their combination (black line) for normal (left panel) and inverted (right panel) neutrino mass ordering. \label{fig:TotMassChns}}
\end{center}
\end{figure}
In Fig.~\ref{fig:TotMassChns} we show the sensitivity to the total mass of the Earth for each different event topology considered, as indicated by the colored lines, as well as for the joint fit accounting for all topologies (black line) assuming the normal ordering (left panel) and inverted ordering (right panel).
We have assumed a total 400~kton-year exposure.
We see that DUNE will be able to measure the Earth's total mass with good precision, 
\begin{equation}
  M = (1\pm0.084)M_\Earth\hspace{2cm}\text{(400~kton-year exposure)}.
\end{equation}
This is much better than the current determination of Earth's mass via the atmospheric oscillation measurement by the Super-Kamiokande collaboration of $M/M_\Earth\simeq 1.04\pm0.21$~\cite{Super-Kamiokande:2017yvm}. 
In the case of IceCube,  only a preference for $M/M_\Earth > 0$ at less than 1$\sigma$ is found~\cite{IceCube:2019dyb}.
Besides, it is interesting to estimate the sensitivity to for a lower exposure.
For example, a 60~kton-year exposure run, corresponding to two far detector modules taking data for three years, would lead to
\begin{equation}
  M = (1^{+0.48}_{-0.43})M_\Earth\hspace{2cm}\text{(60~kton-year exposure)}.
\end{equation}

From this figure we also see that, in the normal ordering case, the sensitivity is driven by charged current (CC) events with 1 outgoing proton and no pions, CC-1p0$\pi$.
This is because at sub-GeV energies, the neutrino-nucleus cross section is dominated by quasi-elastic (QE) transitions in which $\nu_\ell + n\to\ell+p$, where $\ell=e,\mu$.
Final state interactions could change the number of outgoing protons and neutrons, however many times the kicked proton will exit the nucleus without interactions.

For the inverted mass ordering, the situation is different.
While essentially nothing changes for the solar resonance, the atmospheric resonance above a GeV now occurs for antineutrinos. 
In antineutrino scattering, there is a larger chance to knock out neutrons~\cite{Palamara:2016uqu}.
Because of that, CC-0p0$\pi$ events, which have zero protons and zero pions, contribute significantly to the sensitivity.
Moreover, at high Earth masses, the MSW resonance energy is reduced, as can be seen in Eq.~\eqref{eq:energy_msw}.
In this case, the experimental sensitivity comes from the atmospheric resonance, since the solar one happens at too low energy.
Nevertheless, due to the crude directionality of the CC-0p0$\pi$ sample (see Fig.~\ref{fig:Reco2D} bottom, especially bottom-left for low-energy), the sensitivity is reduced, which explains why the $\chi^2$ is suppressed at high masses relative to the normal ordering case.

\begin{figure}
\begin{center}
\includegraphics[width=0.8\linewidth]{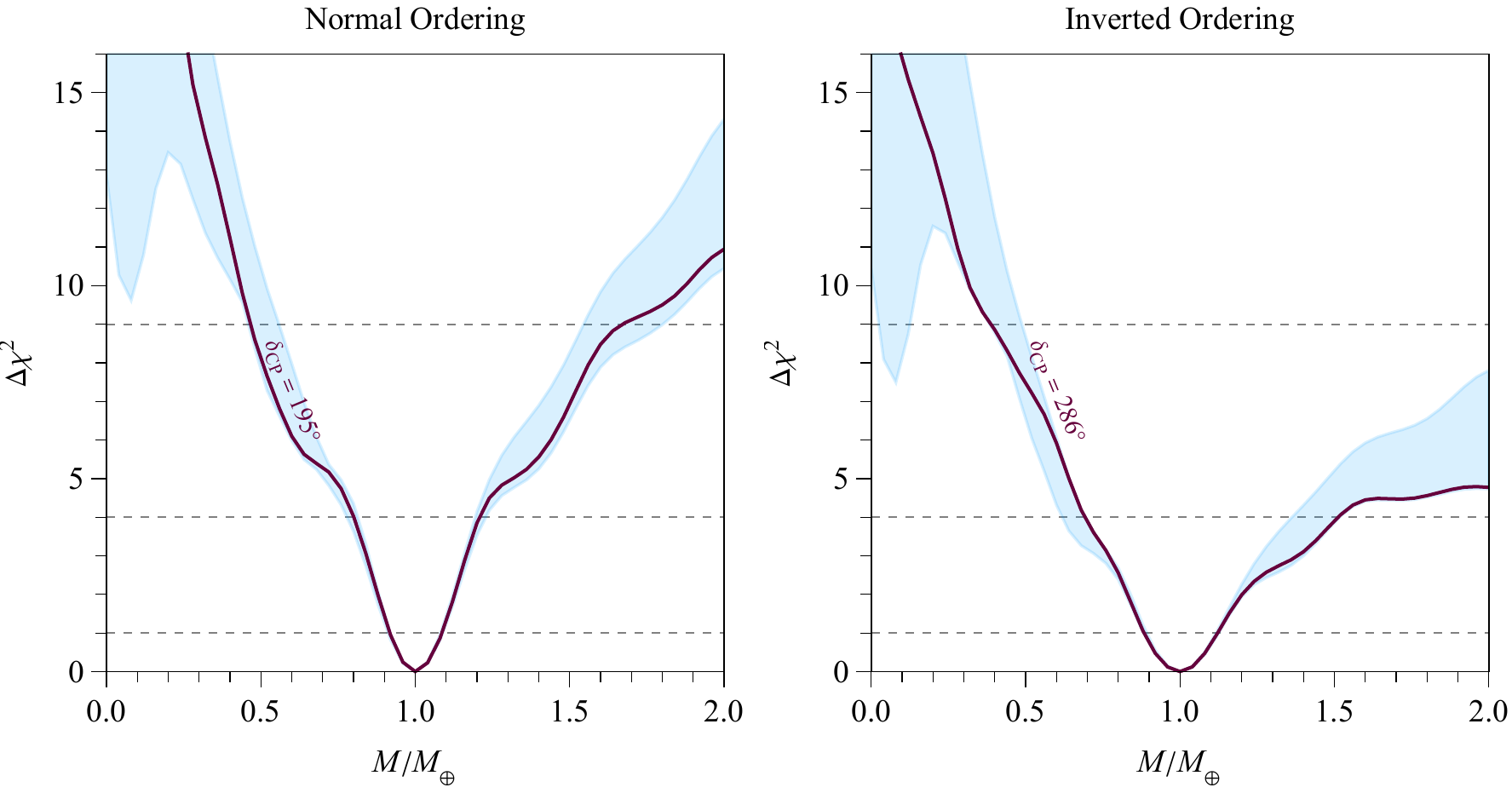}
\caption{DUNE's sensitivity to the total mass of the Earth for the current global best fit value of $\delta_{\rm CP}$ (line) for normal (left panel) and inverted (right panel) neutrino mass ordering. The band encodes the sensitivity varying the true value of $\delta_{CP}$ within $0$ to $2\pi$.
 \label{fig:TotMass}}
\end{center}
\end{figure}

Although most oscillation parameters have been measured well experimentally, the current hint for the $CP$ violating phase strongly depends on a combination of several experimental results.
It is far from clear that the true values of the $CP$ phase are the ones found in the current global fits, and only more data will clarify the situation.
Therefore, one is led to question how much the sensitivity depends on the true value of $\delta_{CP}$.
To answer this question, we show in Fig.~\ref{fig:TotMass} the sensitivity for the best fit values of the $CP$ phase, as well as a band that encodes the minimum and maximum sensitivities for all possible values of $\delta_{CP}$, for a given $M/M_\Earth$.
As we can see, the impact of the $CP$ phase is only large away from $M/M_\Earth=1$, which shows that the our estimate for DUNE's determination of the total mass of the Earth is indeed robust, especially at $1\sigma$. 
We also studied the impact of varying the true values of other oscillation parameters on the measurement of Earth's total mass. 
Within present constraints, we found a negligible variation of the sensitivity to $M/M_\Earth$, in particular near $M/M_\Earth \sim 1$.

\subsection{Measuring the Earth's Matter Density Profile}
\label{subsec:MeasureWithConstraints}
We now proceed to estimate the capability of DUNE in measuring the matter density profile of the Earth.
Again, we consider the 3-layer Earth model introduced in Section~\ref{sec:AtmoNu}.
The radii that separate each layer are related to how seismic waves propagate and get reflected, and are fairly well known by seismological data (see e.g. Ref.~\cite{Geller:2001ix}).
In particular, the liquid core shadows S-waves coming from the opposite side of the planet, which provides an excellent measurement of the core radius.
In view of that, we assume that the radii at which the density profile transitions from the core to the lower mantle and from the lower mantle to the upper mantle are known.
We refer to these radii of interest as $R_{\rm C} = 3480$ km (core $\to$ lower mantle), $R_{\rm LM} = 5700$ km (lower mantle $\to$ upper mantle), and $R_{\rm UM}=R_\Earth = 6371$ km (radius of Earth).

\begin{figure}
\begin{center}
\includegraphics[width=0.5\linewidth]{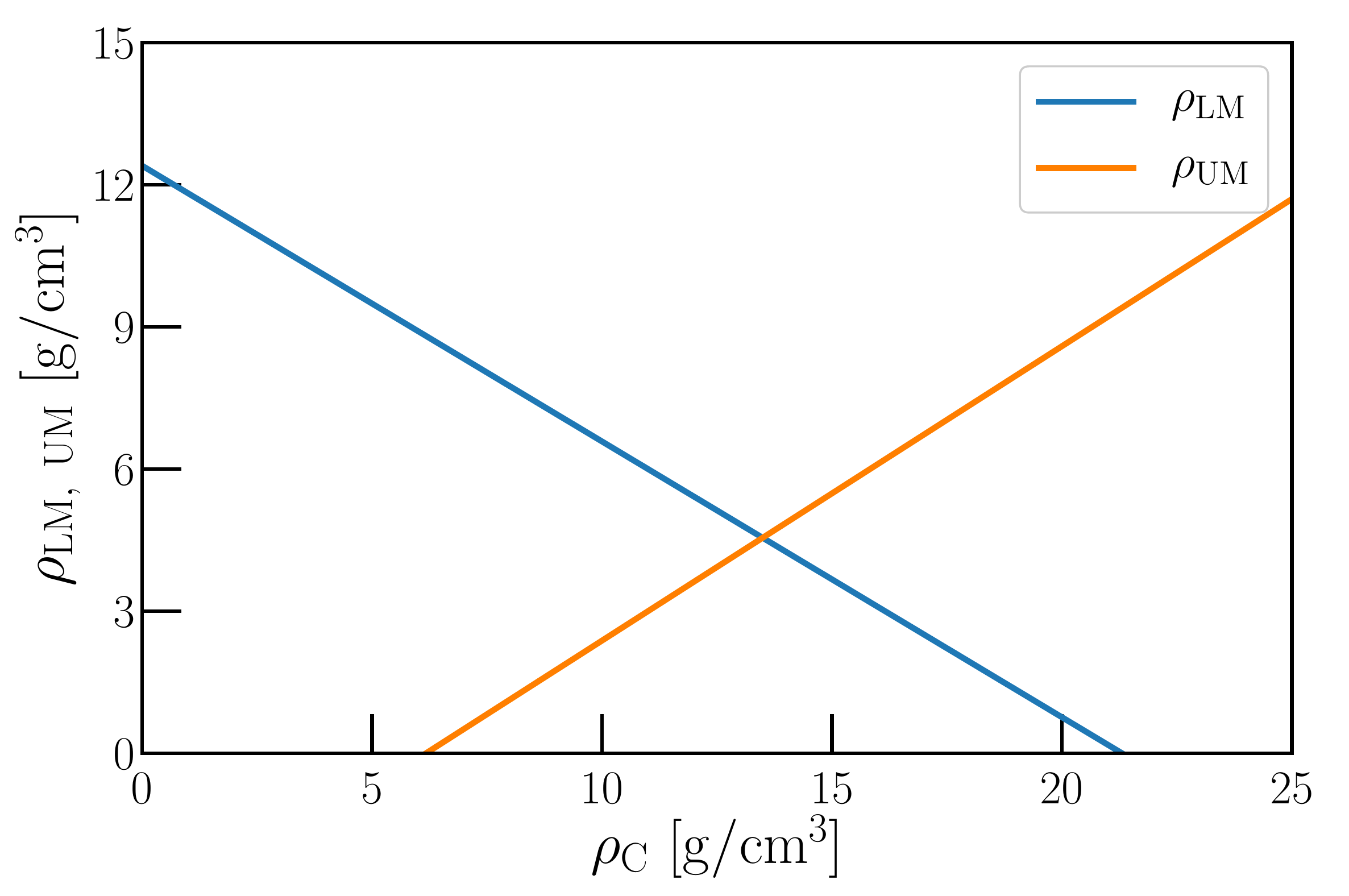}
\caption{Constrained relationship between $\rho_{\rm LM}$ (blue) and $\rho_{\rm UM}$ (orange) as a function of the free input parameter $\rho_{\rm C}$, where we assume that the mass of the Earth and its moment of inertia are known, using the relationships in Eqs.~\eqref{eq:ME} and~\eqref{eq:IE}. See text for further detail.\label{fig:MEIEConstraints}}
\end{center}
\end{figure}

Since we do not expect any sensitivity to the finer details of this density profile, nor the fact that the Earth is not perfectly spherical, we will neglect these higher-order effects. 
The total mass of the Earth $M_\Earth$ can be inferred from its effect on orbiting satellites~\cite{ries1992progress} and independent measurements of the Newtonian gravitational constant $G$~\cite{Rosi:2014kva} ---the current determination being $M_\Earth = 5.9722\times10^{24}$~kg with a $10^{-4}$ relative uncertainty.
In addition, by observing the Earth's precession and nutation, its moment of inertia is inferred to be $I_\Earth=8.01738\times10^{37}$~kg~m$^2$, again with a small $10^{-4}$ relative uncertainty~\cite{williams1994contributions, chen2015consistent}.
Given the precision of such measurements, we will take both mass and moment of inertia of the Earth as constraints on the matter profile, namely,
\begin{align}
M_\Earth &= \frac{4\pi}{3}\left[ \rho_{\rm C} R_{\rm C}^3 + \rho_{\rm LM} \left(R_{\rm LM}^3 - R_{\rm C}^3\right) + \rho_{\rm UM} \left(R_\Earth^3 - R_{\rm LM}^3\right)\right], \label{eq:ME}\\
I_\Earth &= \frac{8\pi}{15}\left[ \rho_{\rm C} R_{\rm C}^5 + \rho_{\rm LM} \left(R_{\rm LM}^5 - R_{\rm C}^5\right) + \rho_{\rm UM} \left(R_\Earth^5 - R_{\rm LM}^5\right)\right].\label{eq:IE}
\end{align}
This procedure provides enough information to require only one independent input. 
We take $\rho_{\rm C}$ to be that independent input, where $\rho_{\rm LM,UM}$ are determined by satisfying Eqs.~\eqref{eq:ME} and~\eqref{eq:IE}. 
The relation imposed by the aforementioned constraints between $\rho_{\rm LM}$ and $\rho_{\rm UM}$ as a function of the input $\rho_{\rm C}$ is shown in Fig.~\ref{fig:MEIEConstraints}.
Due to the constrained relationship by knowing the mass of the Earth and its moment of inertia, as well as our three-layer approximation, we see that this setup limits $6$ g/cm$^3$ $\lesssim \rho_{\rm C} \lesssim 21$ g/cm$^3$.

\begin{figure}
\begin{center}
\includegraphics[width=0.85\linewidth]{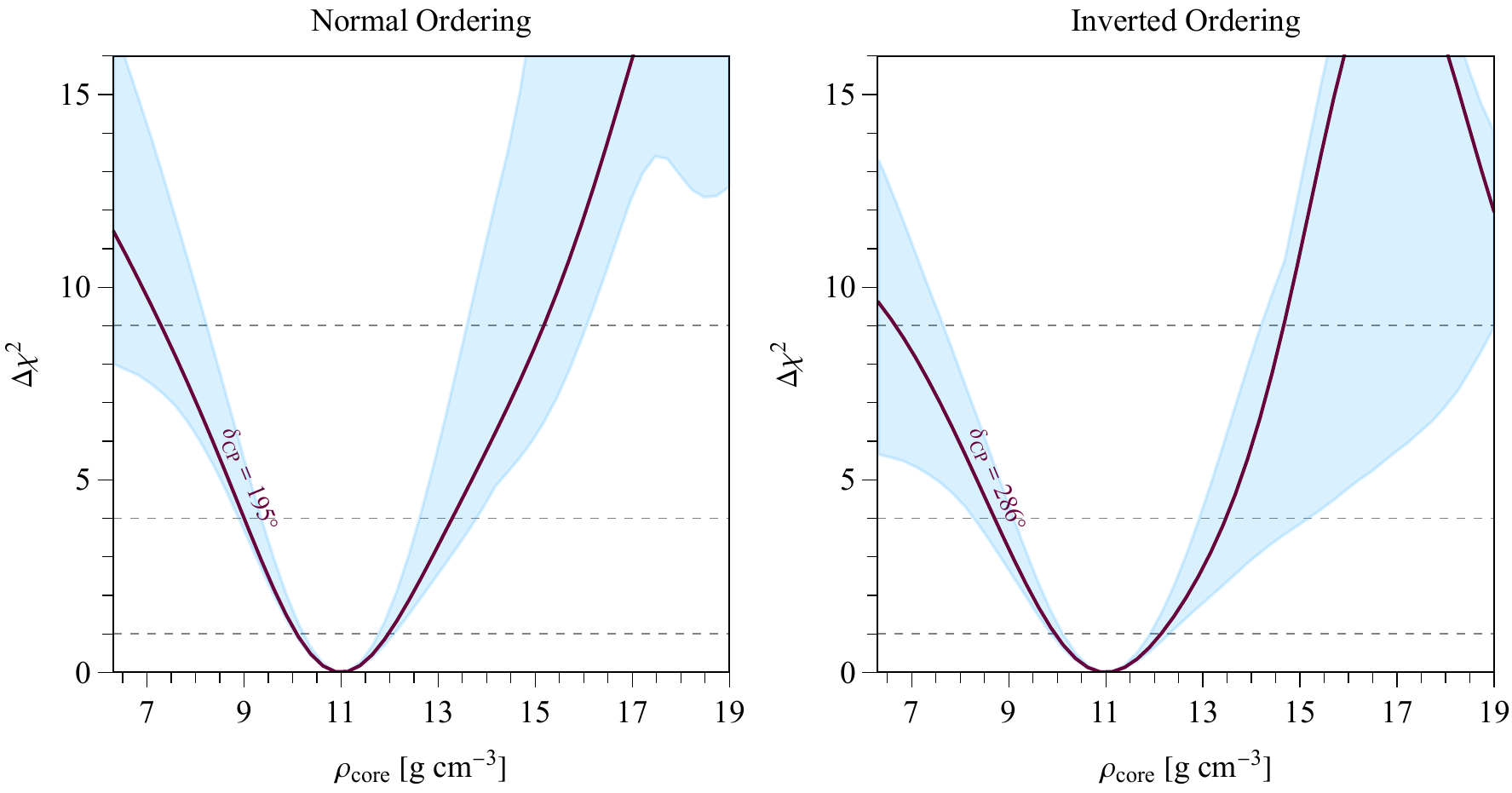}
\caption{DUNE's sensitivity to the core matter density $\rho_{\rm core}$, assuming the density profile to be constrained by the total mass and moment of inertia of the Earth. 
The left (right) panel assumes that neutrinos follow the normal (inverted) mass ordering, where both assume that the ordering is known from the neutrino beam data. 
The blue band encodes different true values of $\delta_{CP}$, while the current global best fit is highlighted (purple lines).\label{fig:MeasRho_MEIE}}
\end{center}
\end{figure}
Allowing $\rho_{\rm C}$ to vary independently, we determine the measurement sensitivity assuming $400$ kt-yr of DUNE operation and present this result in Fig.~\ref{fig:MeasRho_MEIE}.
As before, we perform this simulation for both normal (left panel) and inverted (right panel) neutrino mass orderings. 
We highlight the sensitivity for the current global best fit value of the $CP$ phase (line).
The blue bands envelop the sensitivity estimates as we vary the true value of $\delta_{CP}$.
As before, we fix all other oscillation parameters as their measurements from other current and DUNE-contemporary experiments will be significantly more powerful than what atmospheric neutrinos can provide.

While the $3\sigma$ measurement range of $\rho_{\rm C}$ varies depending on the input parameters (as optimistic as $\rho_{\rm C} \in [7, 14]$ g/cm$^3$ and as pessimistic as $\rho_{\rm C} \in [6, 17]$ g/cm$^3$), we see in Fig.~\ref{fig:MeasRho_MEIE} that the $1\sigma$ extraction is fairly independent of the truth assumption.
Since all three densities are constrained by Earth's mass and moment of inertia measurements, our results indicate that, under these assumptions DUNE will be able to provide the following measurements for a 400~kton-year exposure:
\begin{subequations}
\begin{align}
  &\rho_{\rm C} = 11.0 \times(1^{+0.088}_{-0.083})~{\rm g}/{\rm cm}^3         &\text{(core, 400~kton-year)},&\\
  &\rho_{\rm LM} = 5.11\times(1^{+0.12}_{-0.13})~{\rm g}/{\rm cm}^3        &\text{(lower mantle, 400~kton-year)},&\\
  &\rho_{\rm UM} = 3.15 \times(1^{+0.22}_{-0.20})~{\rm g}/{\rm cm}^3      &\text{(upper mantle, 400~kton-year)}.&
\end{align}
\end{subequations}
For a lower, 60~kton-year exposure, we find that DUNE can measure
\begin{subequations}
\begin{align}
  &\rho_{\rm C} = 11.0 \times(1^{+0.30}_{-0.25})~{\rm g}/{\rm cm}^3         &\text{(core, 60~kton-year)},&\\
  &\rho_{\rm LM} = 5.11\times(1^{+0.35}_{-0.42})~{\rm g}/{\rm cm}^3        &\text{(lower mantle, 60~kton-year)},&\\
  &\rho_{\rm UM} = 3.15 \times(1^{+0.71}_{-0.60})~{\rm g}/{\rm cm}^3      &\text{(upper mantle, 60~kton-year)}.&
\end{align}
\end{subequations}

The 400~kton-year sensitivity is particularly relevant for the core chemical composition.
It is currently believed that the core is composed mainly of iron~\cite{Dziewonski:1981xy}, whose average proton-to-neutron ratio is $p/n\simeq 0.87$, or equivalently $Z/A\simeq0.47$. 
As discussed before, the matter effect on neutrino oscillations is sourced by the electron number density, which should be equal to the proton number density for an electrically neutral object.
Therefore, a determination of the core's proton number density with better than $\sim10\%$ sensitivity using neutrino oscillations could in principle contribute to the understanding of the core's chemical composition.

To put things in perspective, we can compare DUNE's $\sim8.5\%$ relative sensitivity to Hyper-Kamiokande's.
From Ref.~\cite{Hyper-Kamiokande:2018ofw}, Hyper-Kamiokande will be able to measure the $Z/A$ ratio of the core with a relative 6.2\% precision after 10~Mton-year exposure, which would correspond to about 27 years of data taking with two 187~kton modules or 53 years with only one of those modules. 
Given that DUNE's 400~kton-year exposure would correspond to 10 years of data taking with all four 10~kton modules, it may seem surprising that DUNE can compete with a much larger detector on measuring the Earth's matter profile.
The reason for this relies on the ability to measure sub-GeV atmospheric neutrinos, which provide a much high event rate compared to the higher energy flux component: DUNE is expected to measure about 15k neutrino events with reconstructed energy below 1 GeV, versus 5k with energies above 1 GeV, for 400~kton-year exposure.

\begin{figure}[t]
\begin{center}
\includegraphics[width=\textwidth]{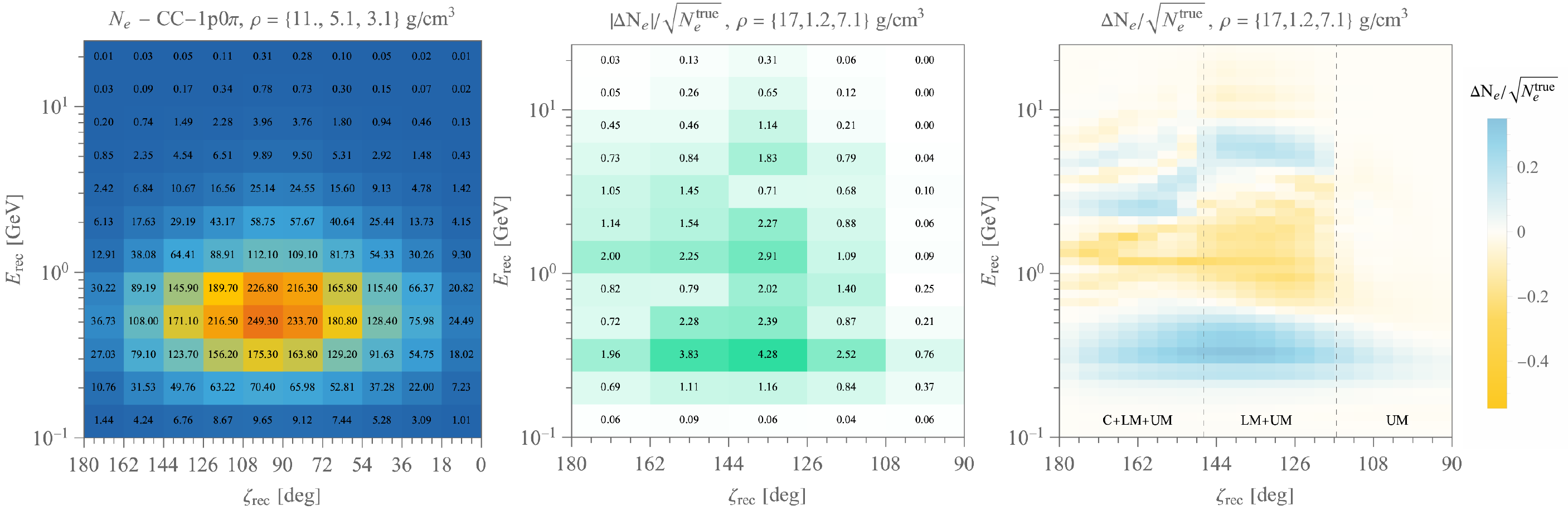}
\caption{\emph{Left}: Number of charged current electron neutrino events with one proton and no pions in the final state(CC-1p0$\pi$), taking into account cross section and detector effects, as a function of the reconstructed neutrino energy and zenith angle. 
\emph{Middle}: Statistical power of CC-1p0$\pi$ sample, see text for details.
\emph{Right}: Statistical power of CC-1p0$\pi$ sample for the $40\times60$ binning used in our simulations; the darkest yellow and blue correspond to $\Delta N_e/\sqrt{N_e^{\rm true}}= 0.35, -0.25$, respectively.
 \label{fig:Evse1p}} 
\end{center}
\end{figure}
Let us discuss in more detail where DUNE's sensitivity comes from.
If on one hand the sub-GeV sample has more statistics than the few-GeV sample, the role of energy and angular resolutions on the final experimental sensitivity is not obvious, see Fig.~\ref{fig:Reco2D}.
To address this, we show in Fig.~\ref{fig:Evse1p} a set of three panels related to electron-neutrino events, accounting for the neutrino-nucleus interaction physics and detector effects described in Sec.~\ref{sec:DUNEDetails}.
The left panel shows the total number of charged current $\nu_e+\overline\nu_e$ events with one proton and no pions (CC-1p-0$\pi$) as a function of the reconstructed zenith angle and neutrino energy for the nominal Earth matter density profile.
To avoid a cluttered figure, we have combined bins to show a $10\times10$ grid.
We can clearly see that the highest statistics comes from neutrinos with 0.4-1~GeV energies and zenith angles such that they cross the lower mantle.

In the middle panel, we show the combined statistical power of each bin for a representative matter profile $\rho=\{17,\,1.2,\,7.1\}~{\rm g/cm^{3}}$ which is ruled out at more than 3$\sigma$ as can be see in Fig.~\ref{fig:MeasRho_MEIE}.
We define the statistical power as the difference of events between the two matter profile assumptions, $\Delta N_e$ divided by the statistical uncertainty $\sqrt{N_e^{\rm true}}$ for each bin.
We calculate that with the original $40\times60$ grid, and sum up the absolute values  $|\Delta N_e|/\sqrt{N_e^{\rm true}}$ in the relevant bins to show the $10\times10$ grid of Fig.~\ref{fig:Evse1p}.
Although systematic uncertainties are taken into account when evaluating the final sensitivity, we do now show them here.
We can see that most of the statistical power comes from neutrinos which cross the core and lower mantle in their trajectories.
Sub-GeV neutrinos do provide most of the statistical power, though there is a non-negligible contribution from the few-GeV sample.

In the right panel of Fig.~\ref{fig:Evse1p}, we show the statistical power in the $40\times60$ grid.
The blue regions indicate downward variations on the number of events while yellow refers to upward.
The darker blue and yellow bins correspond to $\Delta N_e/\sqrt{N_e^{\rm true}}= 0.35, -0.55$, respectively, values for the statistical power.
Although the individual power in each bin is small, one can clearly see that for this representative matter profile there are large regions presenting events consistently above or below the assumed true matter profile case.

\begin{figure}[t]
\begin{center}
\includegraphics[width=\linewidth]{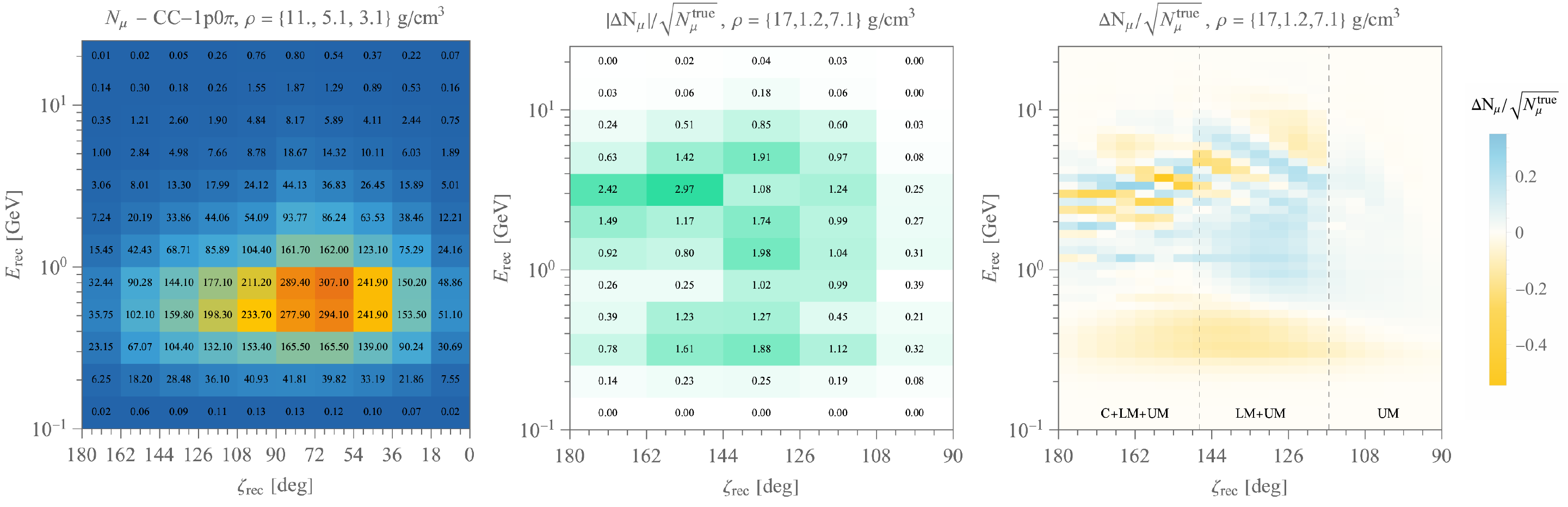}
\caption{\emph{Left}: Number of charged current muon neutrino events with one proton and no pions in the final state (CC-1p0$\pi$), taking into account cross section and detector effects, as a function of the reconstructed neutrino energy and zenith angle. 
\emph{Middle}: Statistical power of CC-1p0$\pi$ sample, see text for details.
\emph{Right}: Statistical power of CC-1p0$\pi$ sample for the $40\times60$ binning used in our simulations; the darkest yellow and blue correspond to $\Delta N_\mu/\sqrt{N_\mu^{\rm true}}= 0.28, -0.55$, respectively.
\label{fig:Evsmu1p}} 
\end{center}
\end{figure}
Last, we highlight that Fig.~\ref{fig:Evse1p} only includes CC-1p-0$\pi$ events for electron neutrinos. 
While the contribution of other topologies to DUNE's sensitivity is typically subleading, unless the mass ordering is inverted, see e.g. Fig.~\ref{fig:TotMassChns}; muon neutrino events do provide significant statistical power to this analysis. 
We present in Fig.~\ref{fig:Evsmu1p} the same panels as in Fig.~\ref{fig:Evse1p}, but now for $\nu_\mu+\overline\nu_\mu$ CC-1p-0$\pi$ events.
Relative to electron events this sample profits more from the atmospheric resonance at few GeV, as can be seen by comparing the middle panels of Figs.~\ref{fig:Evse1p} and \ref{fig:Evsmu1p}.
These two sets of panels show that the atmospheric neutrino sample at DUNE can indeed provide us an oscillogram of Earth, culminating on a world-leading measurement of the Earth matter profile.

\subsection{Measurements without Constraints}
\label{subsec:MeasureRhos}
We now proceed to our last analysis, the measurement of Earth's matter profile, in the 3-layer simplified model of the Earth, without taking into account the measurements of the total mass and moment of inertia of the Earth.
While it may seem unreasonable to disregard both measurements, this exercise will show us that DUNE atmospheric neutrino sample alone can still teach us something about the Earth.

\begin{figure}
\begin{center}
\includegraphics[width=0.8\linewidth]{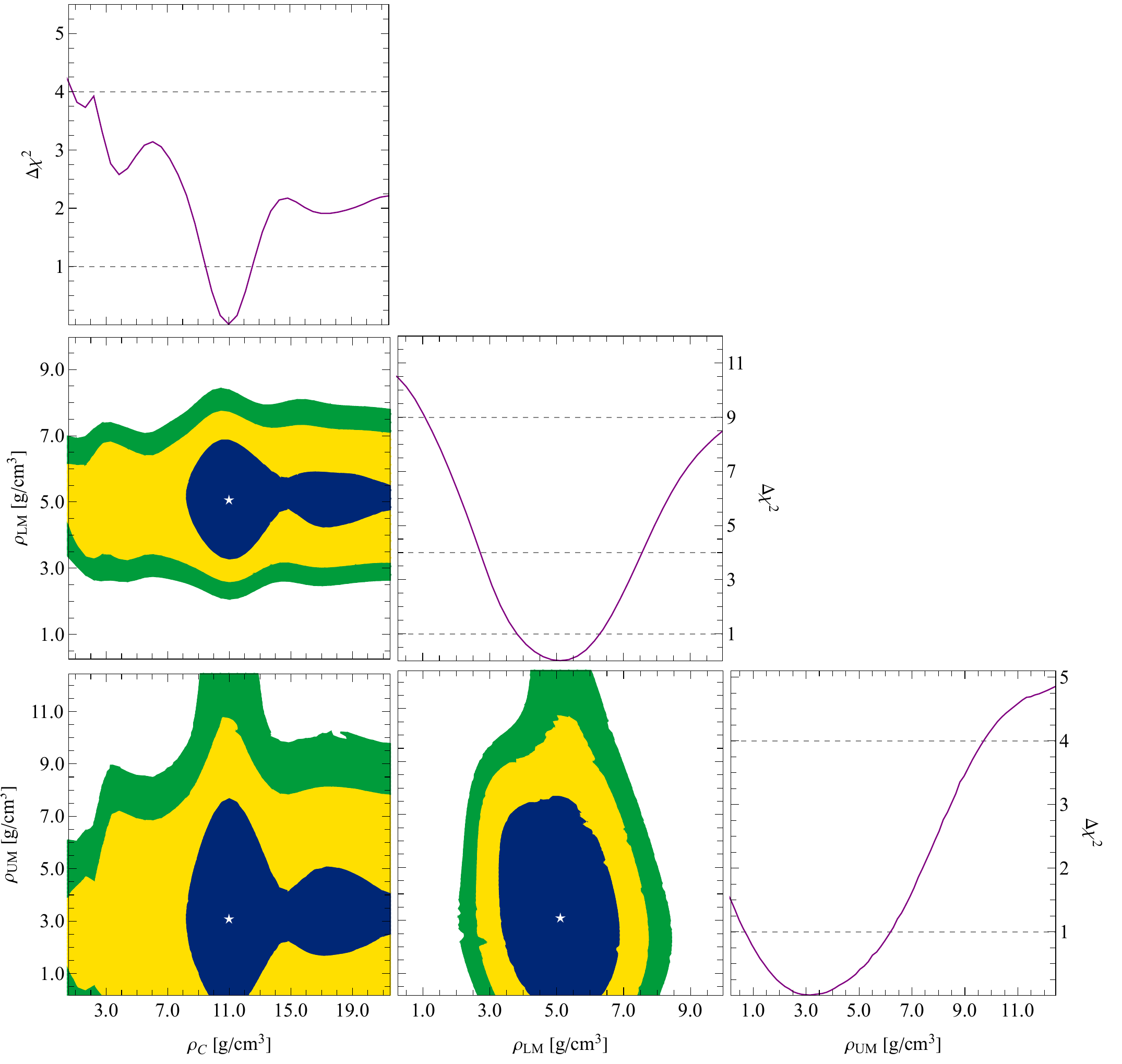}
\caption{DUNE's sensitivity to the core, lower mantle and upper mantle densities without considering constraints from total Earth mass and moment of inertia measurements. The colored areas correspond to 1$\sigma$ (blue), 2$\sigma$ (yellow), and 3$\sigma$ (green) allowed regions for 2 degrees of freedom, that is $\Delta \chi^2=2.3,\,6.2,\,11.8$, respectively. \label{fig:three_layers_no_constraints}}
\end{center}
\end{figure}
For this analysis, we proceed very much as in the previous one, just dropping the constraints \eqref{eq:ME} and \eqref{eq:IE}.
In Fig.~\ref{fig:three_layers_no_constraints} we show the 1$\sigma$ (blue), 2$\sigma$ (yellow), 3$\sigma$ (green) allowed regions in all planes with the core, lower mantle and upper mantle densities, as well as the one-dimensional $\chi^2$ projection.
We can clearly see that DUNE is mostly sensitive to the inner core density, due to a combination of large changes in the oscillation probability with high statistics, see Figs.~\ref{fig:Evse1p} and \ref{fig:Evsmu1p}.
The determination of the core density at high significance, in particular, becomes difficult, although DUNE can still reject zero core density at over 2$\sigma$.
Regardless, the 1$\sigma$ uncertainties on the densities were found to be 
\begin{subequations}
\begin{align}
  &\rho_{\rm C} = 11.0 \times(1\pm 0.14)~{\rm g}/{\rm cm}^3         &\text{(core)},&\\
  &\rho_{\rm LM} = 5.11\times(1^{+0.23}_{-0.25})~{\rm g}/{\rm cm}^3        &\text{(lower mantle)},&\\
  &\rho_{\rm UM} = 3.15 \times(1^{+0.97}_{-0.78})~{\rm g}/{\rm cm}^3      &\text{(upper mantle)}.&
\end{align}
\end{subequations}
For a 60~kton-year exposure and without external constraints, DUNE can only provide a meaningful measurement of the  lower mantle density, namely  $\rho_{\rm LM} = 5.11\times(1^{+0.69}_{-0.62})~{\rm g}/{\rm cm}^3$.

\begin{figure}
\begin{center}
\includegraphics[width=1\linewidth]{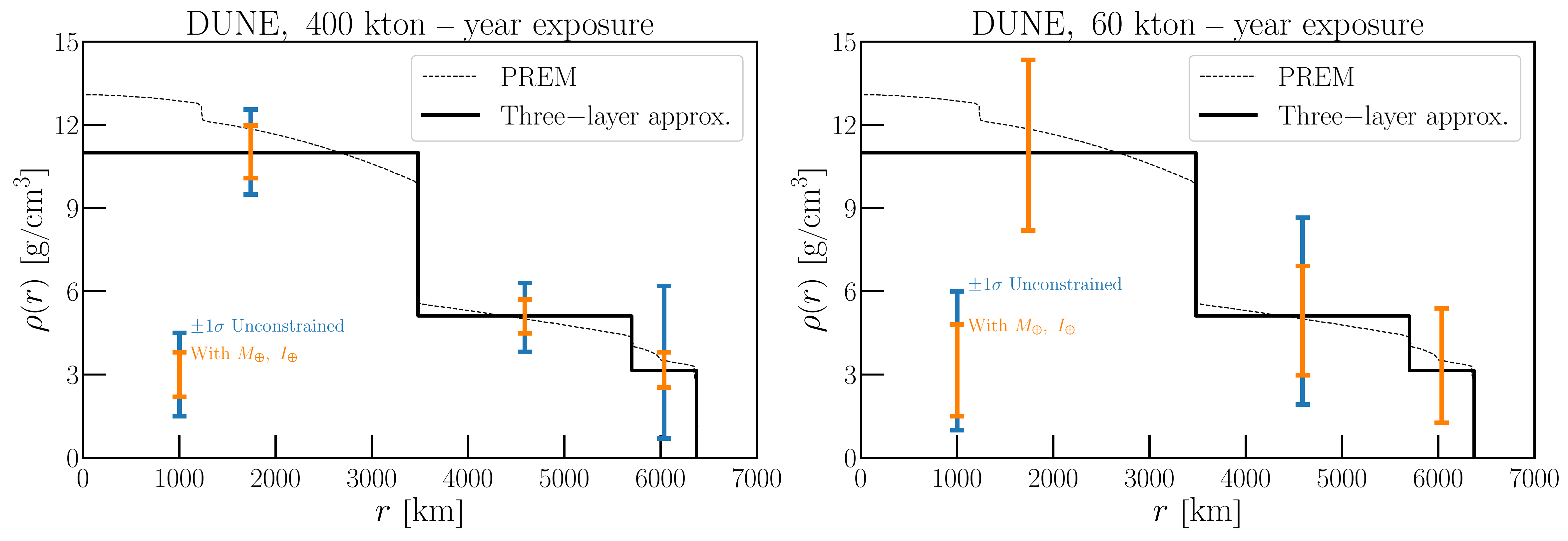}
\caption{DUNE's sensitivity to the Earth matter profile as a function of the radius for the constrained analysis (orange data points), accounting for Earth's total mass and moment of inertia measurements, as well as the unconstrained analysis (blue data points). The solid black line corresponds to our 3-layer Earth model, while the PREM profile is shown for reference as a dashed line. The left panel corresponds to 400~kton-year exposure, while the right panel corresponds to 60~kton-year exposure. Note that the unconstrained analysis with lower exposure does not lead to meaningful determinations of the core and upper mantle densities, and thus we do not display the corresponding blue data points in the right panel.\label{fig:RhovsR}}
\end{center}
\end{figure}

To summarize the results of this and the previous subsection, we show in Fig.~\ref{fig:RhovsR} the 1$\sigma$ uncertainties on DUNE's determination of the matter profile as a function of the radius of the Earth, for 400~kton-year exposure (left) and 60~kton-year exposure (right).
The solid black line corresponds to our 3-layer Earth model, while the dashed line is the PREM density.
Sensitivities are show for the constrained analysis (orange data points), accounting for the total Earth mass and moment of inertia measurements, and for the unconstrained one (blue data points).
Note that in the lower exposure case, there are no meaningful determinations of the core and upper mantle densities, so we omit the corresponding data points.

\section{Conclusions}
\label{sec:Conclusions}
In this paper, we have estimated the sensitivity of the future DUNE experiment to the matter profile of the Earth.
Accounting for measurements of the total mass and moment of inertia of the Earth, we have found that DUNE will be able to measure the core, lower mantle and upper mantle densities with a $8.6\%, 12.3\%, 21\%$ precision, respectively.
Without these external constraints, the precision is degraded to $14\%, 24.3\%, 87.5\%$, respectively. 
When the shape of this profile is assumed to be known, DUNE will be able to determine the total mass of the Earth at the 8.4\% level.
To estimate these sensitivities realistically, we have simulated neutrino-argon interactions using \texttt{NuWro}, a state-of-the-art neutrino event generator; we have considered detector effects in reconstruct individual particles at the event level; and we have accounted for several systematic uncertainties on the atmospheric flux prediction.

We have shown that the reason for DUNE's excellent sensitivity relies on the capability of liquid argon time project chambers to reconstruct event topologies, in particular low energy protons.
This allows the experiment to leverage the large flux of  sub-GeV atmospheric neutrinos, with a decent reconstruction of both their energy and direction.
DUNE can thus leverage the rich phenomenology of MSW and parametric resonances present in the atmospheric neutrino sample to extract information regarding Earth's matter profile.
In fact, DUNE observation of \emph{both} solar and atmospheric matter resonances will be unique among current and future neutrino experiments.
Finally, we have provided a pedagogical description of the physics of both MSW and parametric resonances.


\begin{acknowledgments}
\noindent We would like to thank Stephen Parke for valuable discussions. Fermilab is operated by the Fermi Research Alliance, LLC under contract No. DE-AC02-07CH11359 with the United States Department of Energy. This project has received support from the European Union’s Horizon 2020 research and innovation programme under the Marie Skłodowska-Curie grant agreement No 860881-HIDDeN.

\end{acknowledgments}

\bibliographystyle{JHEP}
\bibliography{refs}

\end{document}